# Sequential Multi-objective Multi-agent Reinforcement Learning Approach for Predictive Maintenance

Yan Chen [a], Cheng Liu [a, b, ]*

[a] Department of Systems Engineering, City University of Hong Kong, Hong Kong, China

[b] Centre for Intelligent Multidimensional Data Analysis, City University of Hong Kong, Hong Kong, China

*Abstract*—Existing predictive maintenance (PdM) methods typically focus solely on whether to replace system components without considering the costs incurred by inspection. However, a well-considered approach should be able to minimize Remaining Useful Life (RUL) at engine replacement while maximizing inspection interval. To achieve this, multi-agent reinforcement learning (MARL) can be introduced. However, due to the sequential and mutually constraining nature of these 2 objectives, conventional MARL is not applicable. Therefore, this paper introduces a novel framework and develops a Sequential Multi-objective Multi-agent Proximal Policy Optimization (SMOMA-PPO) algorithm. Furthermore, to provide comprehensive and effective degradation information to RL agents, we also employed Gated Recurrent Unit, quantile regression, and probability distribution fitting to develop a GRU-based RUL Prediction (GRP) model. Experiments demonstrate that the GRP method significantly improves the accuracy of RUL predictions in the later stages of system operation compared to existing methods. When incorporating its output into SMOMA-PPO, we achieve at least a 15% reduction in average RUL without unscheduled replacements (UR), nearly a 10% increase in inspection interval, and an overall decrease in maintenance costs. Importantly, our approach offers a new perspective for addressing multi-objective maintenance planning with sequential constraints, effectively enhancing system reliability and reducing maintenance expenses.

## 1. Introduction

Engagement in PdM for aircraft engines plays a pivotal role in averting UR, scenarios where engines endure until they have reached the end of their operational lifespan [1]. At the core of estimating RUL and orchestrating PdM for aircraft, a wealth of signal data derived from numerous embedded sensors throughout the aircraft should serve as the foundation [2].

State-of-the-art in PdM: From Traditional Reliability Models to Data-Driven Approaches

However, in recent years, despite extensive exploration of PdM across various intricate facilities beyond turbofan engines alone, many maintenance planning models tend to simplify assumptions regarding system or structural degradation [3] [4] [5] [6]. These studies often not fully utilized the wealth of sensor data gathered from realistic systems. Processes such as the Gamma process, Wiener process, non-homogeneous Poisson process, or Markov process are commonly presumed to govern degradation in PdM research [7]. With the advancement of data-driven technologies, there exists a significant opportunity to utilize real data for analyzing degradation and extracting crucial information related to system RUL [8] [9], which can be leveraged for PdM [10]. However, only a few studies have effectively integrated data-driven approaches for RUL prognostics into predictive maintenance strategies [11]. In a recent study, researchers built a probabilistic RUL prognostics model using CNN for turbofan engines [12], optimizing maintenance planning and showcasing substantial cost reductions compared to traditional methods, highlighting the benefits of data-driven based PdM. Additionally, a method for estimating RUL of Unmanned Aircraft Systems (UAS) using vibration data was developed, which calculates RUL upon exceeding a specified threshold [13], demonstrating its efficacy in predicting system degradation and PdM in UAS operations. Another study presents a method for PdM of industrial robots in intelligent manufacturing [14], combining data and knowledge to predict faults. By utilizing deep learning and knowledge graphs, the approach can automatically formulate PdM strategies.

---

\* Corresponding author.
*E-mail address*: cliu647@cityu.edu.hk

Uncertainty Quantification in RUL Prediction for Robust Maintenance Strategies

Nevertheless, existing investigations often simplify RUL predictions to a point estimation. However, in real-world scenarios, RUL predictions are influenced by various factors such as sensor noise, operational variability, and the complex nature of equipment aging, all of which introduce significant uncertainty. This uncertainty plays a critical role in maintenance decision-making, as ignoring it can lead to suboptimal strategies. Therefore, quantifying the uncertainty of RUL in models is essential, as it not only enhances the model's adaptability to real-world data but also improves the robustness and reliability of decision-making processes. Bayesian learning is a common method for estimating the uncertainty of neural network outputs. For instance, an integrated prognostic-driven dynamic PdM framework for industrial systems has connected degradation features with RULs using Bayesian deep learning [15]. By adjusting maintenance decisions based on predictive RUL distributions, operational constraints are efficiently managed and compared against benchmark policies using turbofan engine data. However, the computational complexity of Bayesian networks is significant. On the other hand, Quantile Regression (QR) has emerged as a promising non-parametric probabilistic prediction method that focuses on forecasting specific quantiles within the forecast distribution, ensuring robustness by avoiding distributional assumptions and providing accurate probabilistic forecasts with precise intervals. For example, researchers leveraged QR based failure prediction method to develop a PdM strategy [16], demonstrating its effectiveness and flexibility. Moreover, a paper proposed a QR-based model to enhance RUL predictions for rolling bearings by addressing uncertainty [17], which shows high accuracy and reliability, indicating improved performance in cross-domain prediction tasks. Therefore, in this study, we leverage QR to estimate the probabilistic distribution of RUL, thereby quantifying the associated uncertainty, and offering richer information to support the development of maintenance strategies.

Beyond Thresholds: Adaptive PdM with Deep Reinforcement Learning

Despite incorporating probabilistic RUL prognostics into maintenance planning, many approaches often rely on fixed degradation thresholds for PdM decision. In this article [18], researchers introduced a novel approach involving training a new model to assess the probability of the predicted RUL falling below a predefined threshold. However, the essence still lies in using fixed thresholds [19], limiting the applicability in real world PdM tasks. Moreover, this reliance within the PdM framework can exhibit instability as data complexity increases. To handle this, a new research has introduced Deep Reinforcement Learning (DRL) for adaptive PdM [20], eliminating the need for fixed thresholds in determining maintenance schedules [21]. In [22], authors discuss using RL for industrial PdM, who proposed a novel approach that combines probabilistic modeling and DRL, demonstrating superior performance in a turbofan engine case study with enhanced interpretability. Furthermore, recently, researchers proposed integrating data-driven probabilistic RUL prognostics into PdM using DRL with Monte Carlo dropout [23]. Also demonstrated on turbofan engines, the approach reduces total maintenance costs and prevents 95.6% of unscheduled maintenance, offering a comprehensive roadmap from sensor data to PdM.

Towards Balanced PdM: Multi-Objective Optimization with DRL

Another problem is, for whether DRL-based or reliant on fixed thresholds, most PdM approaches predominantly center on deciding whether to initiate replacement action in each inspection and pay little attention to the cost caused by checking. But conducting checks after every operational cycle proves economically inefficient, especially during the initial stages of operation when engine degradation is minimal [20]. Consequently, in an adaptive PdM framework, it is vital to simultaneously consider multiple objectives, necessitating the integration of multi-objective optimization into DRL for more balanced and efficient decision-making [24]. Research on MARL has commenced in various domains, showcasing the remarkable capability of DRL in addressing multi-objective PdM tasks.

To address the aforementioned limitations of existing methods and inspired by these advancements, a new framework is proposed in this paper within the context of turbofan engine PdM tasks, aiming to bridge the gap in applying MARL to data-driven PdM. We have 2 main objectives - the first being to reduce the RUL at the time of replacement for maximizing engine utilization, and the second being to minimize the frequency of engine inspections by extending the intervals between inspections as much as possible. Therefore, we introduce 2 RL agents to achieve these objectives. However, due to the sequential nature of these 2 objectives, we cannot apply typical MARL algorithms [25]. To address this, we designed SMOMA-PPO algorithm, inspired by PPO [26], providing a novel approach to solving such problems. On the other hand, inspired by the good performance of Gated Recurrent Unit (GRU) on RUL prediction tasks [27] and in order to provide effective information for RL algorithms to make decisions, QR was utilized to construct the GRP model to obtain a probability distribution of estimated RUL. Compared to previous methods, we achieve efficient and stable Probabilistic Regression, and achieve higher accuracy in the later stages of engine operations. The focus on the later stages is due to the need for PdM task to take replacement actions when the RUL is small, requiring higher precision in last cycles. Finally, our experimental results demonstrate that our method significantly reduces the average RUL at replacement time, maximizes inspection intervals, and substantially decreases overall cost. Our contributions are summarized as follows:

1. Introduction of a new probabilistic RUL prediction method utilizing QR, demonstrating significantly higher prediction accuracy in the later stages of a system's lifecycle compared to existing approaches.
2. Innovative utilization of probability distribution functions to fit the results of QR from the RUL prognostic model, leading to favorable outcomes through the calculation of cumulative probability values of RUL for constructing RL observation environments.
3. Development of a novel multi-agent reinforcement learning algorithm addressing multi-objective optimization problems with temporal dependencies, enabling minimized RUL engine replacements and substantial inspection cost reductions, offering a new data-driven solution for temporally constrained problems.

## 2. Problem Statement

PdM of turbofan engines is a critical yet complex task, demanding precise and adaptive strategies to ensure operational reliability while minimizing maintenance costs. Despite notable advancements in recent research, several critical challenges remain unresolved, limiting the

practical effectiveness of current approaches:

1. RUL prediction serves as the foundation for maintenance planning. However, existing methods often simplify RUL predictions to deterministic point estimates, neglecting the inherent uncertainties present in real-world scenarios. This lack of probabilistic insight weakens the robustness and reliability of maintenance decisions.
2. Modern turbofan engines generate abundant sensor data. Yet, many existing approaches fail to fully exploit this wealth of information, relying instead on oversimplified degradation assumptions or traditional statistical models. This results in suboptimal feature extraction from degradation signals, limiting the ability to capture subtle indicators of impending failures.
3. Common PdM strategies rely on predefined RUL thresholds to trigger maintenance actions. While straightforward, this approach lacks adaptability to dynamic operating conditions and often becomes unstable as data complexity increases, reducing its real-world applicability.
4. Most PdM frameworks overlook the economic inefficiencies associated with frequent inspections. In practice, conducting inspections after every operational cycle is not only costly but also unnecessary. Furthermore, achieving dual objectives—reducing the RUL at replacement to maximize engine utilization while simultaneously minimizing inspection frequency to reduce costs—has been rarely studied due to its inherent complexity. These objectives are further complicated by their temporal dependencies, as decisions made to achieve one goal often impact the timing and feasibility of the other, requiring a sequential resolution strategy that adapts over time.

These challenges underscore the urgent need for a comprehensive, data-driven framework capable of addressing the uncertainty in RUL predictions, fully utilizing sensor data, and dynamically balancing interdependent objectives with temporal constraints. To tackle these issues, this study proposes a novel approach that leverages QR for probabilistic RUL estimation and integrates it with a MARL framework. By bridging these critical gaps, the proposed framework aims to significantly improve the adaptability, robustness, and cost-efficiency of PdM strategies for turbofan engines.

## 3. Methodology

To achieve efficient and reliable PdM for turbofan engines, this study introduces a novel framework that integrates probabilistic RUL prediction with a sequential multi-objective multi-agent RL approach. The framework addresses two critical objectives: (1) reducing the RUL at replacement to maximize engine utilization, and (2) minimizing inspection frequency to reduce operational costs. These objectives are achieved through the combination of advanced RUL prediction techniques and a MARL framework specifically designed to handle sequential, interdependent decision-making tasks.

As illustrated in Fig. 1, the framework begins with the extraction of essential information from raw sensor data using a GRP model (top-right corner of Fig. 1). This model utilizes QR to estimate the probability distribution of the RUL, allowing the system to quantify prediction uncertainty and generate richer information for decision-making. The resulting RUL distribution is then transformed into cumulative probability values for specific RUL intervals (e.g., RUL = 0 to 10 cycles), as shown in the green box on the left of Fig. 1. These cumulative probabilities compactly encode both the likelihood of failure and the uncertainty, serving as the input state representation for the RL agents. The bottom section of Fig. 1 illustrates the SMOMA-RL framework, where two RL agents collaborate to achieve the dual objectives:

1. Agent 1 determines whether to replace the engine during the current cycle, aiming to minimize RUL at replacement while ensuring safety.
2. Agent 2 predicts the optimal time for the next inspection, maximizing the interval

between inspections while maintaining reliability.

A key innovation of this framework lies in the sequential dependency between the 2 agents: the inspection intervals determined by Agent 2 affect the observations available to Agent 1, while the replacement action of Agent 1 resets the inspection process for Agent 2. This interdependence requires a careful design of the RL algorithm. To address this, the proposed SMOMA-PPO algorithm adapts Proximal Policy Optimization (PPO) to handle multi-agent, multi-objective tasks with temporal dependencies.

The following sections detail the technical components of the framework, starting with the GRP model (Section 3.1) and progressing to the SMOMA-PPO algorithm (Section 3.2). Together, these components enable a robust and adaptive solution for real-world PdM tasks.

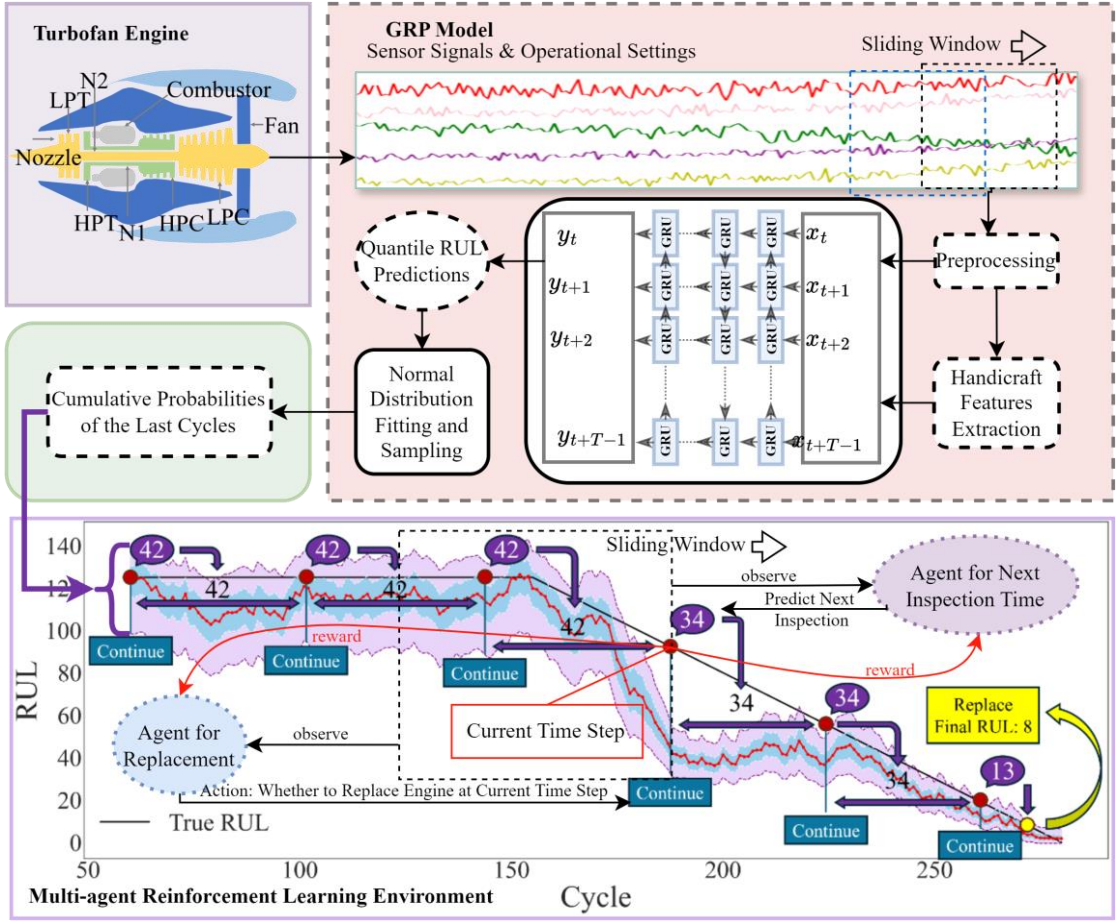

**Fig. 1.** Overall process diagram of this study. Sensor signals were obtained from the Turbofan Engine, then key features were extracted using a deep learning model based on GRU. Subsequently, quantile regression and function fitting were employed to estimate the RUL range, which is utilized as the observation state for MARL agents for executing PdM actions.

3.1. Gru-based Probabilistic RUL Prediction Model

3.1.1. Point Estimation & Probabilistic RUL prediction

At the timestamp t, the real RUL $y_t$ is as Eq. (1).

$$y_t = f(x_t; \theta) + \varepsilon_t \tag{1}$$

where $f(x_t; \theta)$ is a deep network with input $x_t$ and learnable parameters $\theta$. $\varepsilon_t$ is noise. $x_t$ can contain multiple historical observations $z_t, \ldots z_{t-l+1}$, where $l$ is the sliding window length.

Usually, noise $\varepsilon_t \sim \mathcal{N}(0, \sigma^2)$ was assumed to follow Normal distribution. Thus, the parameters $\theta$ of deep network $f(\cdot)$ can be estimated by solving Eq. (2).

$$\theta^* = \text{argmin}_\theta \frac{1}{N} \sum_{i=1}^{N} (y_i - |f(x_i; \theta|))^2 \tag{2}$$

where $\{x_i, y_i\}_{i=1}^{N}$ are training samples. After obtaining the estimated $\theta^*$, $\hat{y}_t = f(x_t; \theta^*)$ can be calculated as the predicted RUL at $t$. $\hat{y}_t$ is said to be a point estimate of $y_t$. At time $t$, make probabilistic RUL prediction as Eq. (3).

$$\mathbb{P}(y_t \mid x_t) = \mathbb{P}(y_t \mid z_t, \ldots z_{t-l+2}, z_{t-l+1}) \tag{3}$$

After conditional distribution, point estimate $\hat{y}_t$ can be obtained as an expectation as Eq. (4).

$$\hat{y}_t = \mathbb{E}[y_t \mid x_t] \tag{4}$$

Thus, the main task is designing a deep network $f(x_t; \theta)$ incorporating historical observations and RULs to model a conditional distribution. However, traditional point estimation methods fail to capture the uncertainty inherent in the prediction process as we discussed in the introduction. To quantify the uncertainty in predictions, our work proposed a probabilistic RUL prediction model generating a posterior estimated RUL distribution using QR. QR for distribution estimation: Multiple predicted RULs at different quantile levels can be obtained by QR method. Given $\hat{y}_i^q$ outputted by $f(x_i; \theta)$ at a quantile level $q \in [0, 1]$, and the real RUL $y_i$, the QR loss is as Eq. (5).

$$L_q(y_i, \hat{y}_i^q) = q(y_i - \hat{y}_i^q)^+ + (1 - q)(\hat{y}_i^q - y_i)^+ \tag{5}$$

where $(y_i)^+ = \max(0, y_i)$. Thus, the final optimization problem using QR loss is as Eq. (6).

$$\theta^* = \text{argmin}_\theta \frac{1}{N} \sum_{i=1}^{N} \sum_{m=1}^{M} L_{q_m}(y_i, \hat{y}_i^{q_m}) \tag{6}$$

where $q_m \in \{q_1, \ldots, q_M\}$ are multiple quantile levels. In our TE PM task, $q_m \in \{0.1, 0.3, 0.5, 0.7, 0.9\}$. The quantile $\hat{y}_t^{0.5}$ can be regarded as the predicted RUL and $[\hat{y}_t^{0.9}, \hat{y}_t^{0.1}]$ is the interval estimation with 80% confidence. This approach overcomes the limitations of traditional point estimation methods by enabling the quantification of uncertainties, allowing the system to dynamically account for risks during decision-making. In the subsequent context, we will provide a detailed explanation of the GRU-based RUL prediction network.

3.1.2. Gated Recurrent Unit

The GRU [28, 29] module was invented for statistical machine translation. It integrates the forget gate and input gate of LSTM [30, 31] to create a unified update gate, encompassing both the cellular state and hidden state. The computation of the hidden unit activation $r_t$ at time step $t$ is shown as Eq. (7).

$$r_t = \sigma(W_r h_{t-1} + U_r x_t) \tag{7}$$

where $W_r$, $U_r$ and $\sigma$ are weight matrices and logistic sigmoid. $\tilde{h}_t$ will then be generated by $r_t$ with a tanh layer as Eq. (8).

$$\tilde{h}_t = \tanh(W(r_t^* h_{t-1}) + U x_t) \tag{8}$$

GRU create a $z_t$ to replace the remember gate and forget gate in LSTM as Eq. (9).

$$z_t = \sigma(W_z h_{t-1} + U_z x_t) \tag{9}$$

At last, the hidden state value is updated as Eq. (10).

$$h_t = (1 - z_t) * h_{t-1} + z_t * \tilde{h}_t \tag{10}$$

Compared to LSTM, GRU offers a more compact and efficient architecture with fewer variables, resulting in improved performance across various tasks.

### 3.1.3. Self-Attention Mechanism

Attention mechanism [32, 33] captures the concept of human tendency to focus on specific regions of an image during recognition, which means that different regions of an image can be assigned with distinct weights. In RUL prediction, it can be leveraged to assign varying weights to different features at different time steps. The features learned by the GRU network for a sample is denoted as $H = \{h_1, h_2, ..., h_d\}^T$. Here, $h_i \in R_n$ and $n$ denotes the number of sequential steps. Then, the importance of different sequential steps for the ith input $h_i$ is as Eq. (11), where $W$ is weight and $b$ is bias.

$$s_i = \Phi(W^T h_i + b) \tag{11}$$

The score function $\Phi(\cdot)$ is an activation function within neural networks. Subsequently, $s_i$ will be normalized as Eq. (12).

$$a_i = softmax(s_i) = \frac{exp(s_i)}{\sum_i exp(s_i)} \tag{12}$$

The final output $O$ of self-attention operation is as Eq. (13). $A = \{a_1, a_2, ..., a_d\}$ and $\otimes$ is element-wise multiplication.

$$O = H \otimes A \tag{13}$$

### 3.1.4. GRU-based RUL Prediction Model

The GRP model integrates advanced deep learning techniques with handcrafted feature extraction to enhance the accuracy and reliability of RUL prediction. As shown in Fig. 2, raw sensor data is first fed into a GRU network to extract sequential features that capture temporal degradation patterns. These features are further refined using a self-attention mechanism to assign attention weights to different time steps, emphasizing the most relevant information for prediction. Meanwhile, handcrafted features, i.e., the mean and trend coefficients of signals from sliding windows, are extracted to provide additional degradation-related insights. These handcrafted features have been shown to improve RUL prediction performance [34].

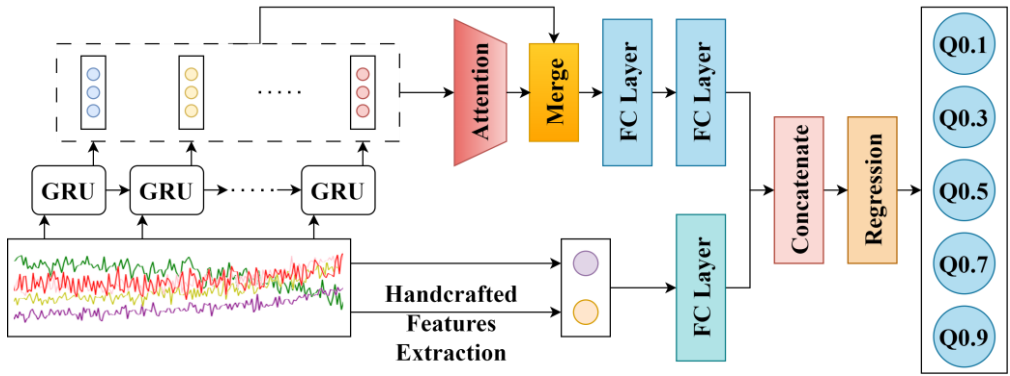

**Fig. 2.** GRU-based probabilistic RUL prediction model, the outputs are the quartiles of predicted RULs.

To maximize the utilization of all available information, a feature fusion framework is proposed, combining the sequential features learned by the GRU network with the handcrafted features. The fused feature set is then processed by fully connected layers (FC) and a regression layer to predict RUL quantiles, specifically the 10%, 30%, 50%, 70%, and 90% percentiles. These quantiles provide a probabilistic representation of the RUL, enabling the model to capture both point estimates and uncertainty. To address potential overfitting and improve model generalization, dropout [35] is applied after the FC layers.

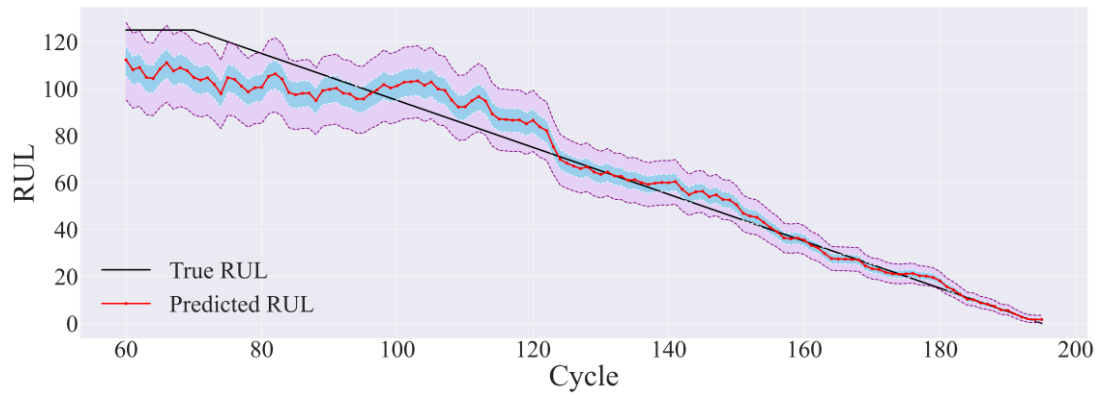

**Fig. 3.** Model prediction of engine 53 in FD001, where the purple region represents the 10%-90% confidence interval of RUL, and the blue area represents the 30%-70% interval.

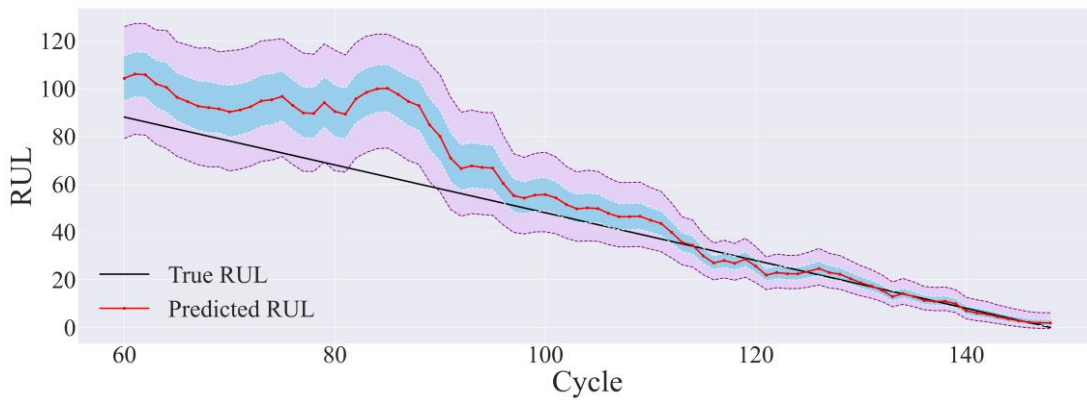

**Fig. 4.** The model prediction of engine 180 in FD002.

In this study, we utilized the "training files" of 4 subsets (FD001 to FD004) from the C-MAPSS [36] dataset as our research objectives. Fig. 3 and Fig. 4 illustrate the performance of the GRP model on 2 randomly selected engines: engine 53 from FD001 and engine 180 from FD002. The figures show the predicted RUL quantiles (10%, 30%, 50%, 70%, and 90%) at each operational cycle, plotted alongside the actual RUL. The red line represents the 50% quantile (the median prediction), while the blue region indicates the range between the 30% and 70% quantiles, and the pink region represents the 10% to 90% quantile range. As observed, the GRP model's predictions become increasingly accurate as the engine approaches the later stages of its lifecycle, with the predicted quantiles closely enveloping the actual RUL. This demonstrates the model's ability to effectively capture the uncertainty in RUL predictions while maintaining high predictive accuracy, particularly in the critical final cycles.

However, while the GRP model outputs discrete RUL quantiles that describe specific probability levels, these quantiles alone do not provide a continuous representation of the RUL uncertainty, which is critical for further analysis and decision-making. A complete probability distribution enables a more detailed understanding of equipment degradation and allows for a comprehensive characterization of uncertainty. Moreover, from the full distribution, multiple cumulative probabilities can be calculated, which are essential as inputs for PdM decision-making. Therefore, it is necessary to fit these quantiles into a continuous distribution. Based on the observed symmetry in the quantiles around the median (50th percentile) and the higher density near the center, and supported by the experiments in section 5.3, a Normal distribution is selected as the best fit compared to other candidates such as Laplace and Cauchy distributions.

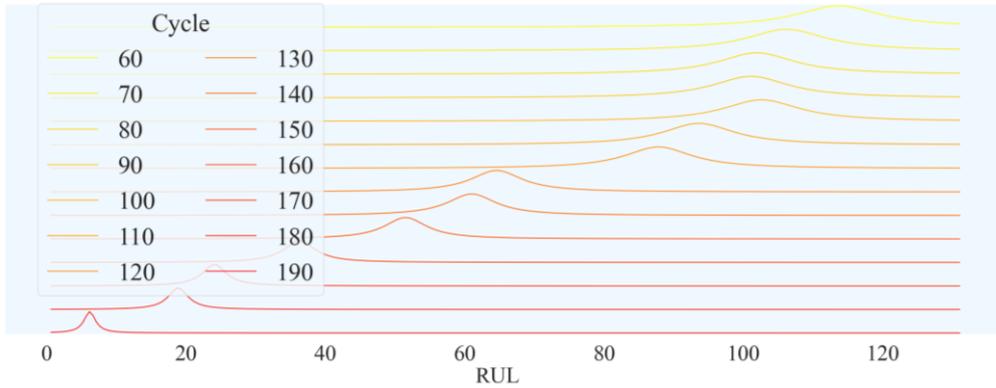

**Fig. 5.** The RUL probability distribution of the engine at different cycles was achieved by Normal distribution fitting.

Fig. 5 demonstrates the fitted RUL distributions for various cycles of engine 53 from the FD001 dataset. The peak of the distribution shifts leftward over time, accurately capturing the progressive decline in the most probable RUL as the engine approaches failure. These distributions are critical for transforming RUL predictions into actionable insights.

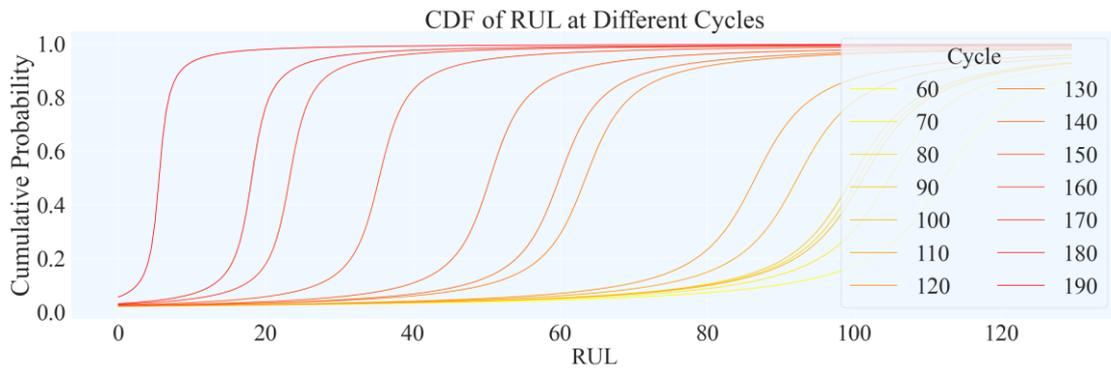

**Fig. 6.** The curve of RUL cumulative probability distribution of the engine at different cycles by Normal distribution fitting.

For downstream PdM task, the continuous RUL probability distributions are further transformed into cumulative probability values. These cumulative probabilities represent the likelihood of the engine's RUL falling within specific intervals (e.g., RUL = 1–10 cycles) and provide a compact, uncertainty-aware state representation that is suitable for RL agents. Fig. 6 illustrates the cumulative probability curves for different cycles, showing the evolution of failure likelihood as the engine degrades. This transformation not only captures prediction uncertainty but also encodes critical information for downstream decision-making.

3.2. Sequential Multi-objective Multi-agent Reinforcement Learning Framework

For the PdM challenge of aircraft engine, we aim to simultaneously achieve multiple objectives. The first goal is for the AI system to replace the engine before it reaches its maximum lifespan. Here, the term "before" suggests maximizing the engine's utilization, with the engine's RUL being minimized at the replacement moment. The second objective is to minimize the frequency of inspection. In practical terms, each inspection incurs human and material costs, thus conducting checks after every flight is highly inefficient. A good way to address such multi-objective problems is to employ MARL, assigning each agent the responsibility of optimizing a specific objective. For the majority of current MARL applications, researchers typically model multiple similar RL agents. These agents act simultaneously upon observing the environment, where similarity refers to these agents usually sharing the same structure of policy network and same action space. In this research, we have 2 objectives. The first goal is to minimize the RUL of the engine when replacing it, and the second is to reduce engine inspections as much as possible. These objectives correspond to 2 actions. The first action involves the replacement of engine based on current sensor data, ideally when the engine's RUL is minimal. The second one involves the agent determining the next inspection time also based on current sensor data, with the aim of maximizing the inspection interval. These actions have a clear sequential order, where the execution of the second action affects the observations of the first agent on the next inspection point. The execution of the first action determines the need for the next inspection, thereby determining the effectiveness of the second agent's action. Furthermore, the action spaces of these two agents also exhibit significant differences. These characteristics deviate significantly from existing typical MARL algorithms. In order to address this challenge, we have innovatively proposed SMOMA-PPO, which is capable of executing such complex sequential tasks.

3.2.1. MDP Modeling for PdM Planning

In this study, each objective of PdM will be achieved by an RL agent. We first formulate the PdM problem as a Markov Decision Process (MDP). Subsequently, we will summarize the proposed algorithm. MDP is a mathematical framework used to model decision making problems in a stochastic environment. It provides a formal way to represent a sequential decision-making process where an agent interacts with an environment. The objective in an MDP is to find a policy, which is a mapping from states to actions, that maximizes the cumulative expected reward over time. The MDP is represented as a tuple to describe the sequential decision-making process of an agent. Specifically, the MDP can be defined as $\mathcal{M} = (S, A_r, A_p, P, R)$, where $S, A_r, A_p, P$ and $R$ denote the state space, action space of agent 1/object 1, i.e., whether to replace the turbofan engine, action space of agent 2/object 2, i.e., the time steps for next inspection from now on, state transition probability, and reward function, respectively.

1) State Space $S$: The state space encompasses the entire range of possible states that agents may occupy. In our scenario, despite substantial differences in the action spaces of the two agents, their observation spaces remain identical. This implies that local and global observation states are indistinguishable, eliminating the need to differentiate between the observation spaces of distinct agents. In this MDP, the agent receives a state $s_t$ generated by the GRP model at each time step, which is the cumulative probabilities of the last $D$ cycles calculated from the estimated RUL distribution: $s_t = [p_{1,t}, \dots, p_{D,t}]$, where $p_{k,t}$ for $k \in \{1, \dots, D\}$ is the probability that current RUL is less than $k$ cycles.

2) Action space of agent 1/object 1 $\mathcal{A}_r$: The action space $A_r$ represents the set of all possible actions that the agent 1, targeting object 1, can take in each state. Then, agent 1 takes an action $a_t^r \in \mathcal{A}_r$ based on a policy $\pi_{\theta_1}^r$, which is a mapping function: $\pi_{\theta_1}^r: S \to \mathcal{A}_r$, where $\pi_{\theta_1}^r(s_t)$ was used to denote the action $a_t^r$ performed by the agent 1 at state $s_t$. The policy $\pi_{\theta_1}^r$ can be considered a decision maker, represented by a neural network with parameter $\theta_1$. Given state $s_t$, the agent chooses either to perform a replacement of the engine at cycle $t$, or do nothing as:

$$a_t^r = \begin{cases} 1, & \text{perform replacement at cycle } t, \\ 0, & \text{do nothing.} \end{cases}$$

3) Action space of agent 2/object 2 $\mathcal{A}_p$: Similarly, the action space $\mathcal{A}_p$ represents the set of all possible actions that the agent 2, targeting object 2, can take in a given state. Then, agent 2 takes an action $a_t^p \in \mathcal{A}_r$ based on a policy $\pi_{\theta_2}^p$, which is a mapping function: $\pi_{\theta_2}^p: S \to \mathcal{A}_r$, where $\pi_{\theta_2}^p(s_t)$ was used to denote the action $a_t^p$ performed by the agent 2 at state $s_t$. The policy $\pi_{\theta_2}^r$ can be considered a predictor, represented by a neural network with parameter $\theta_2$. Given state $s_t$, the agent 2 decide the time steps of next engine inspection from current cycle $t$, which was scaled by a pre-defined maxim period $D$: $a_t^p \in \mathbb{N}$ $and$ $0 \leq a_t^p \leq D$.

4) Reward Function $R$: The reward function provides a quantitative measure of the desirability or value associated with being in a particular state and taking particular actions. In order to accomplish efficient and reliable predictive maintenance, we design a reward function that encourages agent 1 to replace the turbofan engine when the RUL at current cycle is as less as possible, but it needs to replace the engine before failure, which corresponding to the UR. We also encourage the agent 2 to predict the inspection gap as large as possible, because we need to save the resource of inspection.

Next, specific reward or penalty rules are formulated for each of the two agents based on their respective objectives.

Objective 1: Replace the turbofan engine when RUL is minimized.

Due to the intricate degradation process of the engine and the influence of various random factors, achieving precise insight into the RUL of the engine during its nonlinear degradation process is an impractical endeavor. Therefore, it is unrealistic to expect the agent to replace the engine exactly at the end of each cycle before it reaches its maximum lifespan. Moreover, in reality, as the engine approaches failure, its state becomes unstable. Hence, ideally, the engine should be replaced within a small gap period before it is on the brink of failure. To establish a rational reward policy that motivates agent 1 to progress towards minimizing the RUL, we introduce two critical reward thresholds, $T_1$ and $T_2$, representing cycle numbers with $T_1 > T_2$. When agent 1 decides to replace the engine, if the current RUL is below $T_1$, the environment will provide a relatively small positive reward by multiplying the current cycle number by a small constant. If the actual RUL at replacement falls between $T_1$ and $T_2$, a larger positive reward is given by multiplying the current cycle number by a larger constant. However, if the RUL at replacement is below $T_2$, a significant negative reward will be assigned by multiplying the current cycle number by a large negative constant as penalty feedback to agent 1. In this study, agent 1 will obtain a reward $r_t^r$ from the environment based on the hidden state $\rho_t$ and action $a_t^r$. The reward $r_t^r$ obtained at decision time point/cycle $t$ is defined for 4 cases considering action $a_t^r$ and the hidden state $\rho_t$ as Eq. (14).

$$r_t^r = \begin{cases} c_1 \times t - c_0 & \text{if } a_t = 1 \text{ and } \rho_t > T_1 \\ c_2 \times t - c_0 & \text{if } a_t = 1 \text{ and } T_2 < \rho_t \leq T_1 \\ -c_3 \times t - c_0 & \text{if } a_t = 1 \text{ and } \rho_t \leq T_2 \\ -c_4 & \text{if } a_t = 0 \text{ and } \rho_t \leq 0. \end{cases} \quad (14)$$

Here, $c_1 \times t$ ($c_1 > 0$) denotes the cost of a scheduled replacement at cycle $t$. $c_0$ is a fixed cost of replacement ($c_0 > 0$). We assume that a too-early replacement is less wanted. Therefore, $c_1$ should be small. Also, we presume $c_2$ is a considerable positive number, i.e., $c_2 > c_1$, since the interval ($T_2$, $T_1$] is where we want replacement to happen, and the closer to $T_2$, the less wasted life caused. $c_3$ denotes the cost of an unwanted (but not unscheduled, if $1 \leq T_2$) replacement. We assume $c_3 > c_2$ since an unwanted replacement is dangerous as it is close to failure. Finally, $c_4 \gg c_3$ is a fixed punishment for unscheduled replacement.

Objective 2: The intervals between detection times should be as long as possible.

To achieve this objective, knowledge of the environmental latent state-the real RUL label at the time of the next inspection is essential for evaluating the efficacy of agent 2's actions. In this context, we still need to utilize the 2 evaluation thresholds defined earlier, $T_1$ and $T_2$, but this time they are used to assess $\rho_{t+1}$ instead of $\rho_t$ as Eq. (15). Once agent 2 determines the next inspection time, the environment can then determine the latent state $\rho_{t+1}$ at the next inspection, representing the actual RUL at that time. If this RUL is less than $T_1$, the environment will provide a relatively small positive reward to agent 2 by multiplying the predictive value by a small constant. If the RUL at the next inspection falls between $T_1$ and $T_2$, the environment will reward agent 2 by multiplying the predictive value by a larger positive constant. However, if the RUL at the next inspection is below $T_2$, agent 2 will be penalized with a significant negative reward by multiplying the predictive value by a large negative constant. This incentivizes agent 2 to provide larger predictions before the RUL reaches $T_1$ and strive to schedule the inspection between $T_1$ and $T_2$.

$$r_t^p = \begin{cases} c_1 \times a_t^p & \text{if } \rho_{t+1} > T_1 \\ c_2 \times a_t^p & \text{if } T_2 < \rho_{t+1} \leq T_1 \\ -c_3 \times a_t^p & \text{if } \rho_{t+1} \leq T_2 \end{cases} \qquad (15)$$

In this study, once agent 2 chooses an action, the hidden state $\rho_{t+1}$ and the observed state $s_{t+1}$ are also updated accordingly. The transition probability $P(\cdot | s_t, a_t^r, a_t^p)$ in this task decides the next state $s_{t+1}$ and is deterministic, which means that the environment transfer from the current state $s_t$ to the next state $s_{t+1}$ according to the order of the sampled states at each epoch. If the engine is replaced at decision step $t$, the next step considers a new engine from the C-MAPSS data set. Otherwise, we further obtain sensor measurements $x_{t+1}$ during the next cycle and update the distribution of the RUL (the next state $s_{t+1}$) by generating new RUL prognostics using the GRP model. When interacting with the environment, the trajectory $\tau$ will be recorded, which is a sequence of interactions from the state $s_0$ to the terminal state $s_m$: $\tau = \{(s_t, a_t^r, a_t^p, r_t)\}_0^m$. It can be used for optimizing 2 agents' policy $\pi_t^r$ and $\pi_t^p$. In summary, the reward function of this PdM task can be formulated as the weighted sum of the aforementioned 2 rewards as Eq. (16).

$$R_t = \beta_1 r_t^r + \beta_2 r_t^p \qquad (16)$$

3.2.2. SMOMA-PPO Algorithm

In this section, we formally present the proposed algorithm for PdM. As illustrated in Fig. 7, the implementation of the algorithm involves training 2 networks for each agent, i.e., the critic network and the actor network, and each agent has different actor and critic. For agent 1, the critic network aims to learn a mapping function $V_{\phi_r}^r$ from the state space $S$ to real values $R_t$. On the other hand, the actor network learns a mapping function $\pi_{\varphi_r}^r$ from the observation space $S$, which is the same as state space in this research, to the action distribution $a_t^r$ for sampling. For agent 2, the critic network also aims to learn a mapping function $V_{\phi_p}^p$ from $S$ to real values $R_t$ and the actor network learns a mapping function $\pi_{\varphi_p}^p$ from $S$, to the action distribution $a_t^p$ for sampling.

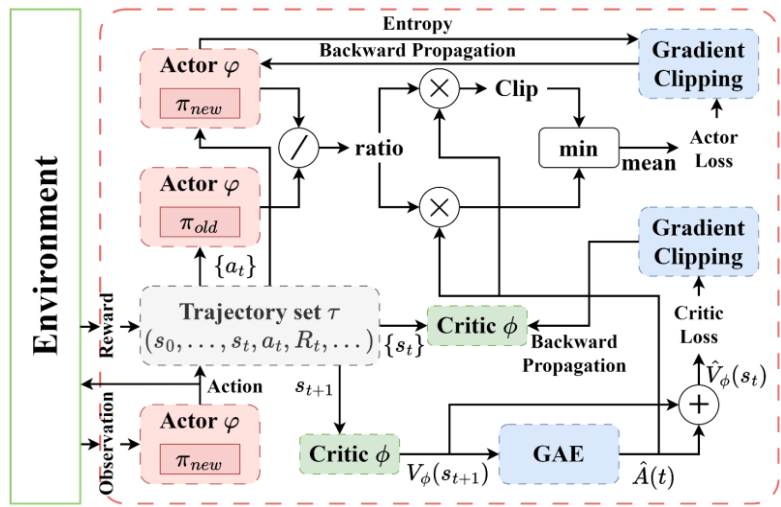

**Fig. 7.** Training process of individual agents in the proposed SMOMA-PPO.

The policy gradient implementation iterates over $T$ time steps to trace trajectory $\tau$, utilizing the gathered samples for subsequent updates. Within the realm of RL, the agent's primary aim is to iteratively refine its policy through engagements with the environment, with the overarching goal of maximizing cumulative rewards. Within the MAPPO algorithm, the advantage function plays a pivotal role in evaluating the benefits of taking a specific action within the current state concerning other available actions, which serves to indicate the expected variance in rewards between actions, guiding the agent towards choices that promise higher returns. In its pursuit of enhancing the precision of advantage function estimation and fostering a more consistent training regime during policy optimization, MAPPO seamlessly integrates the principles of Generalized Advantage Estimation (GAE), as exemplified by: $\hat{A}_t^i = \zeta_t^i + (\gamma\varrho)\zeta_{t+1}^i + \cdots + (\gamma\varrho)^{T-t+1}\zeta_{T-1}^i$, where $i = r, p$, and $\zeta_t^i = R_t + \gamma V_{\phi_i}(s_{t+1}) - V_{\phi_i}(s_t)$. $\gamma$ is the reward discount factor and $\varrho$ is the GAE discount factor. Then, the value function assessed by GAE can be given by Eq. (17).

$$\hat{V}_{\phi_i}(s_t) = \hat{A}_t^i + V_{\phi_i}(s_{t+1}) \tag{17}$$

On this basis, the optimization objective of the actor network is formulated by proximal policy gradient with the goal of maximizing the cumulative reward as Eq. (18).

$$\mathcal{L}(\varphi) = \frac{1}{2T}\left\{\sum_{t=1}^{T}\sum_{i=r,p}\epsilon H[\pi_{\varphi_i}(a_t^i \mid s_t)]\right\}$$

$$+ \frac{1}{2T}\left\{\sum_{t=1}^{T}\sum_{i=r,p}\min[\chi_i(\varphi_i,t)\hat{A}_t^i, clip_{1-\varepsilon}^{1+\varepsilon}\{\chi_i(\varphi_i,t)\}\hat{A}_t^i]\right\} \tag{18}$$

where $min[x, y]$ represents the minimum value between $x$ and $y$. Moreover, $\chi_i(\varphi_i, t) = ([\pi_{\varphi_i}(a_t^i \mid s_t)]/[\pi_{\varphi_i}(a_t^i \mid s_t)])$ and $\varepsilon$ is the PPO clip ratio. Moreover, the entropy $H[\pi_{\varphi_i}(a_t^i|s_t)]$ with coefficient $\epsilon$ can be formulated as Eq. (19).

$$H[\pi_{\varphi_i}] = -P_{\pi_{\varphi_i}(a_t^i|s_t)} log\left(P_{\pi_{\varphi_i}(a_t^i|s_t)}\right) \tag{19}$$

and $P_{\pi_{\varphi_i}}$ is the probability distribution of the strategy. On the other hand, the objective of the critic network is to minimize the mean squared error of the value function as Eq. (20).

$$\mathcal{L}(\phi) = \frac{\sum_{t=1}^{T}\sum_{i=r,p}\left(V_{\phi_i}(s_t) - \tilde{V}_{\phi_i}(s_t)\right)^2}{2T} \tag{20}$$

With the configurations detailed above, the SMOMA-PPO approach can be encapsulated in Algorithm 1. Initially, the algorithm kicks off by orthogonalizing the network parameters for both the actor and critic networks. Throughout the training cycle, two agents gather trajectory segments of length $T$ in each iteration. Each agent, based on its local observation $s_t$, selects an action $a_t^i$ to interact with the environment via the actor network. Subsequently, the agents acquire and store new observations, states, and rewards during this interaction, repeating this loop until reaching time step $T$ to construct a full trajectory. Leveraging the gathered trajectory set, the algorithm employs the GAE method to approximate the advantage function $\hat{A}$. Following this, the optimization procedure for both the actor and critic networks kicks off using the Adam optimization method, incorporating gradient clipping and a decaying learning rate. This training regimen is reiterated for a predetermined number of episodes denoted by $E$.

```
Algorithm 1: SMOMA-PPO
    Initialization: Initialize parameter ϕ for π, φ for V.
        Set learning rate $l_r$.
        for episode = 1 → E do
            set data buffer D = {};
            for n = 1 → batch size do
                Set empty list τ [ ].
                for t = 1 → t = T do
                    for agent i in {r, p} do
                        $P_i(t) = \pi_{\varphi_i}(a_{n,t}^i | s_{n,t})$.
                        $a_{n,t}^i \sim P_i(t)$.
                        $V_i(t) = V_{\phi_i}(s_{n,t})$.
                    end for
                    Execute $a_{n,t}$
                    Obtain R(t)
                    Observe $s_{t+1}$.
                    $\tau += [s_t, a_t, V(t), R(t), s_{t+1}]$.
                end for
                Compute advantage estimate Â by GAE on τ .
                Compute $\tilde{V}_\phi$ with value normalization on τ .
                Split trajectory τ into chunks of length L
                for l = 0,1,…, T//L do
                    $D = D \cup (\tau[l: l + T, \hat{A}[l: l + L], \hat{R}[l: l + L]])$.
                end for
            end for
            for mini-batch k = 1,…, K do
                b ← random mini-batch from D with all agent data
                for each data chunk c in the mini-batch b do
                    Update V from first hidden state in data chunk
                end for
            end for
            Adam update ϕ on L(ϕ) with data b.
            Adam update φ on L(φ) with data b.
        end for
```

The policy network in SMOMA-PPO, also known as actor network, is structured as shown in Table 1 and Table 2, while the critic network structure is illustrated in Table 3. The input dimension for both the actor and critic networks is set to be 50 because the agent utilizes information from the RUL distribution of the previous 5 timesteps for decision-making. For each timestep, we calculate the cumulative probability of RUL ranging from 0 to 10 based on the RUL distribution. Therefore, there are a total of 5×10=50 input features. The critic network structures for agent 1 and agent 2 are the same. However, for Agent 1, which has only 2 actions. Therefore, the output dimension of actor network 1 is 2. Regarding Agent 2, in this study, the output dimension of its actor network is set to be 50 because we assume its predicted inspection time intervals range from integers 1 to 50.

**Table 1.** Actor Network Structure for Agent 1

| Layer | Type | Input Size | Output Size | Activation | Parameter Size |
|---|---|---|---|---|---|
| 1 | Linear | 50 | 128 | Tanh | 6400 |
| 2 | Linear | 128 | 128 | Tanh | 16384 |
| 3 | Linear | 128 | 2 | Tanh | 256 |

**Table 2.** Actor Network Structure for Agent 2

| Layer | Type | Input Size | Output Size | Activation | Parameter Size |
|---|---|---|---|---|---|
| 1 | Linear | 50 | 128 | Tanh | 6400 |
| 2 | Linear | 128 | 128 | Tanh | 16384 |

| 3 | Linear | 128 | 50 | Tanh | 6400 |

**Table 3.** Critic Network Structure for Agent 1 & 2

| Layer | Type | Input Size | Output Size | Activation | Parameter Size |
|---|---|---|---|---|---|
| 1 | Linear | 50 | 256 | Tanh | 12800 |
| 2 | Linear | 256 | 64 | Tanh | 16384 |
| 3 | Linear | 64 | 1 | Tanh | 64 |

5. Experiment

5.1. Data Description

The proposed approach is evaluated using the widely used C-MAPSS dataset [36] in Prognostics and Health Management (PHM) or PdM community, which illustrates the degradation of engines, as depicted in Fig. 1. To monitor the engine's condition, 21 sensors are positioned to measure parameters like temperature, pressure, and speed. The dataset comprises four subsets, each encompassing distinct operational conditions and fault types, as shown in Table 4. In every subset, a training file records sensor data during the run-to-fail experiments for a certain number of engines, while a testing file contains sensor measurements for specific running cycles of another set of engines.

**Table 4.** C-MAPSS Dataset Description.

| Sub-dataset | FD001 | FD002 | FD003 | FD004 |
|---|---|---|---|---|
| Training engines number | 100 | 260 | 100 | 249 |
| Testing engines number | 100 | 259 | 100 | 248 |
| Operational conditions | 1 | 6 | 1 | 6 |
| Fault modes | 1 | 1 | 2 | 2 |

As explained in section III, given our objective of life-long PdM, we focus on utilizing run-to-fail engines. Therefore, the "training file" from C-MAPSS was selected and portioned into a training dataset for model training and a testing dataset for testing. Since different operating conditions can also impact the RUL, we consider the operating settings to be signals for RUL prediction. Consequently, we utilize data from the 21 sensors and 3 columns of operation settings, which were treated as 24 signals.

5.2. Data Preprocessing

The sliding window is widely adopted for data segmentation [37, 38]. For run-to-fail engines, we assign $T$ to represent the total number of running cycles, $s$ to denote the window size, and $p$ to indicate the step size. Each sample will have a size of $s \times n$, where $n$ corresponds to signal types. We have opted for a window size of 60 and a step size of 1 for all subsets, resulting in a size of $s \times n = 60 \times 24$ for every sample. RUL of the $(i + 1)$th sample can be computed as $T - s - i \times p$. It is essential to note that a piece-wise linear RUL is employed, which ensures that if the RUL exceeds the maximum RUL, it is capped at the maximum value of 125. To compare with the performance of related research [23], we adopted the same data partitioning approach, i.e., the training files are divided into two parts: the first 50% is allocated as a training dataset, and the remaining as a testing dataset. The training samples devised by fixed window for four subsets are: 6,959, 19,188, 10,131, and 23,036. (For FD004, the training file contains records for 249 engines, where the first 124 are used for training.) Similarly, the number of testing samples for the four subsets are: 7,772, 19,231, 8,689, and 23,522.

5.3. Performance of GRP and Fitting Distribution Selection

Before evaluating the performance of SMOMA-PPO, we first validate the effectiveness of the GRP model, we conducted a comparative study against several state-of-the-art RUL prediction methods, including CNN-LSTM-SAM [39], GCU-Transformer [40], TCNN-Transformer [41], and RNN-LSTM [42]. Since these methods did not originally employ QR, their regression components were adapted to output quantile predictions for a fair comparison. Specifically, the 50th percentile (median) of each method's outputs was used as the predicted RUL, as it corresponds to the maximum likelihood estimate of the RUL distribution.

The experiments were conducted using FD001. Given the critical importance of accurate RUL predictions during the final stages of engine operation, the Root Mean Square Error (RMSE) [33] was calculated over the last 30 cycles of the engines' lifecycles. Each method was trained and tested 50 times to ensure robustness, and the minimum, maximum, median, mean, and standard deviation of RMSE values were recorded. Additionally, we measured the shortest training time required for convergence to evaluate the computational efficiency of each method.

**Table 5.** Comparison Between Popular RUL Prediction Approaches

| Approach | Minimum | Maximum | Median | Mean | Standard Deviation | Training Time (s) |
|---|---|---|---|---|---|---|
| GRP | 3.9173 | 3.1034 | 3.5465 | 3.5255 | 0.2070 | 32.55 |
| CNN-LSTM-SAM | 4.8418 | 3.1544 | 3.7922 | 3.7963 | 0.3622 | 41.13 |
| GCU-Transformer | 4.9699 | 3.4375 | 4.2633 | 4.2670 | 0.3930 | 86.39 |
| TCNN-Transformer | 4.6465 | 3.9385 | 4.3766 | 4.3726 | 0.1691 | 107.87 |
| RNN-LSTM | 4.9755 | 3.1066 | 4.1581 | 4.1824 | 0.4456 | 113.01 |

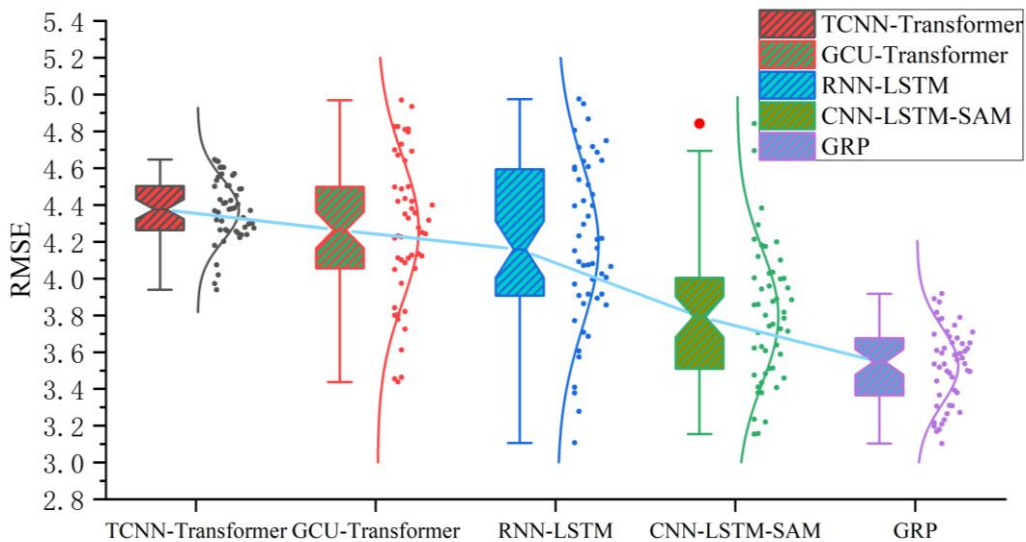

**Fig. 8.** The comparative results of various RUL prediction methods, with the average RMSE values obtained by each method connected by a light blue line. On the right side of each box, a Normal distribution is employed to model the experimental results.

The results, presented in Table 5 and Fig. 8, demonstrate that the proposed GRP model significantly outperformed all other methods in terms of predictive accuracy, robustness, and computational efficiency. The GRP model achieved the lowest mean RMSE of 3.5255 with a standard deviation of 0.2070, reflecting both high accuracy and consistency across runs. This is particularly critical during the final stages of engine operation, where precise RUL predictions are essential for effective maintenance decisions. Fig. 8 visually compares the RMSE distributions from 50 experiments for each method using box plots. The GRP model not only shows a lower median RMSE but also exhibits a narrow interquartile range, indicating greater reliability and small variability. In addition to accuracy, the GRP model is highly efficient, requiring only 32.55 seconds for training, significantly less than GCU-Transformer (86.39 seconds) and TCNN-Transformer (107.87 seconds). This balance between strong predictive performance and computational efficiency highlights the strengths of the GRP model's compact architecture and its well-designed feature fusion framework.

Next, to determine the most suitable probability distribution for fitting the RUL quantiles obtained by GRP, we evaluated three most commonly used symmetric distributions: Normal, Cauchy, and Laplace. The uniform distribution was excluded due to its inability to capture the higher density near the median observed in the quantile predictions. Four engines (engine IDs: 10, 15, 20, and 25) were selected using segmented random sampling from the FD001 testing dataset to ensure representative evaluation and demonstration. For each engine, quantile predictions at 110, 120, 130, and 140 cycles were used to fit the distributions. The 50th percentile was set as the central parameter (mean for Normal, location for Cauchy and Laplace), while gradient descent was employed to optimize the shape parameters (standard deviation for Normal, scale for Cauchy and Laplace). The absolute fitting error was computed for the 10th, 30th, 50th, 70th and 90th percentiles.

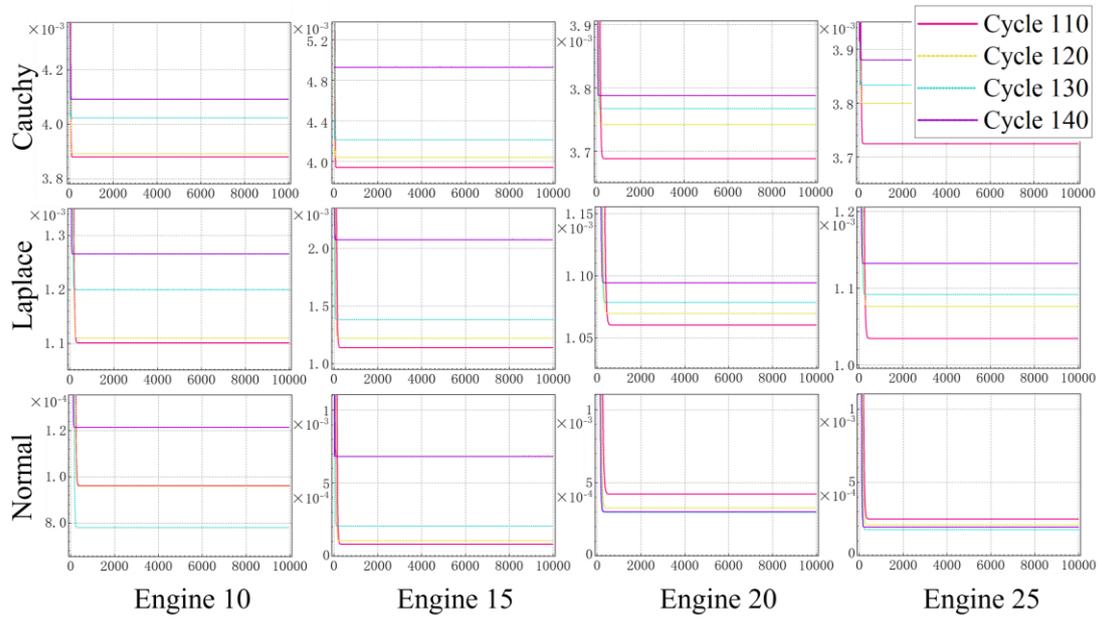

**Fig. 9.** The performances of fitting quantile regression results using different probability distributions.

As shown in Fig. 9, the Normal distribution exhibited the best performance, achieving the lowest absolute error at convergence. The error stabilized at approximately $3\times10^{-4}$, which was an order of magnitude lower than the errors of the Laplace and Cauchy distributions. The Normal distribution also demonstrated consistent behavior across all tested engines and cycles, confirming its reliability for fitting RUL quantiles obtained by GRP. Therefore, the fitted Normal distribution was then used to calculate the cumulative probability values, providing a compact and uncertainty-aware representation of RUL predictions, enabling considerate decision-making in the following PdM task.

5.4. Evaluation Criteria for PdM

In the PdM task, for each subset, the experiment was repeated 10 times. The true RUL of each engine at its final stoppage was recorded in each test, and then the following metrics will be calculated: **M**ean **R**UL (MR); **S**tandard Deviation of **R**ULs (SR); **Ma**ximum **R**UL (MAR); **Mi**nimum **R**UL (MIR); **Med**ian **R**UL (MDR); number of **S**uccessful **Re**placements (SRE): A successful replacement means that the engine was replaced when its RUL was between 0 and 30; **U**nscheduled **R**eplacement-UR: replacement was not performed until RUL=0; **W**asted **Re**placements (WRE): the number of engines that were replaced when their RUL exceeded 30.

## 5.5. PdM Experiment Setup

As shown in Fig. 10, we began with GRP training, and the last 50 cycles of all training engines were used for validation, i.e., the RMSE was calculated after each training. When RMSE no longer decreased, the model was considered to have finished training. Afterward, using the quantile predictions and Normal distribution fitting, we obtained the RUL cumulative probability values of each engine in both training and testing sets for the last ten cycles of engine lifespans. Subsequently, the cumulative probability values will be utilized for DRL training. Finally, we evaluated the DRL model using cumulative probability values generated from testing dataset. DRL training and testing will be conducted 10 times.

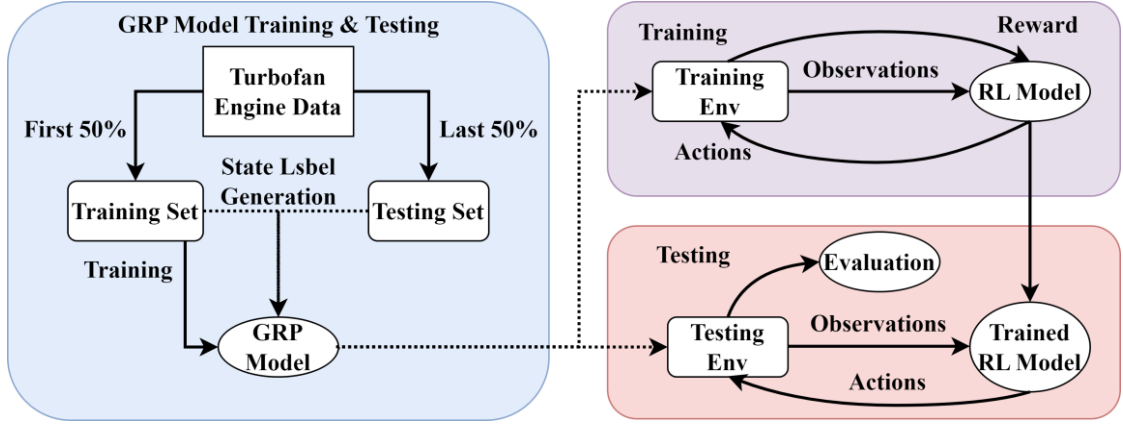

**Fig. 10.** The overall experimental process.

As shown in Fig. 11, we provide a detailed depiction of the training process of SMOMA-PPO by illustrating the evolution of training rewards across five independent random trials. The experiments were performed on FD001. The visualization includes raw reward signals (faint, noisy lines), smoothed curves (lighter solid lines), and their average trends (red dashed line). This comprehensive view highlights how the total rewards (Reward 1 + Reward 2) stabilize around 17,500 seconds, maintaining a range between 1.0 to $1.6 \times 10^6$. Consequently, our training strategy of this research is halting training after 20,000 seconds to ensure robustness.

The individual evolution of Reward 1 and Reward 2 is presented in Fig. 12 and Fig. 13, offering deeper insights into the collaborative dynamics between the 2 agents. Reward 1, approximately 10 times larger than Reward 2, dominates the training process due to its critical role in ensuring system safety and avoiding failures. Failures, particularly unscheduled replacements, incur significant penalties, making Reward 1 the primary driver of total rewards. As the replacement agent (Agent 1) learns to optimize replacement timing and prevent failures, Reward 1 steadily increases and eventually stabilizes at a high level in the later stages of training.

In contrast, Reward 2, associated with inspection scheduling, exhibits a distinct trend. During the early phases of training, the inspection agent (Agent 2) explores maximizing inspection gaps to improve efficiency, resulting in higher Reward 2 values. However, larger inspection gaps also elevate the risk of failures, which negatively impacts Reward 1. Over time, Agent 2 adjusts its policy by slightly reducing inspection gaps, sacrificing some Reward 2 to improve the overall system performance. This trade-off showcases the complementary roles of the two agents: Reward 1 prioritizes safety and reliability by avoiding failures, while Reward 2 enhances cost-efficiency by optimizing inspection intervals within safe operational limits. The interplay between these rewards demonstrates how the two agents collaboratively achieve a robust and balanced maintenance strategy, with Agent 1 focusing on failure prevention and Agent 2 fine-tuning inspection policies to enhance overall system performance.

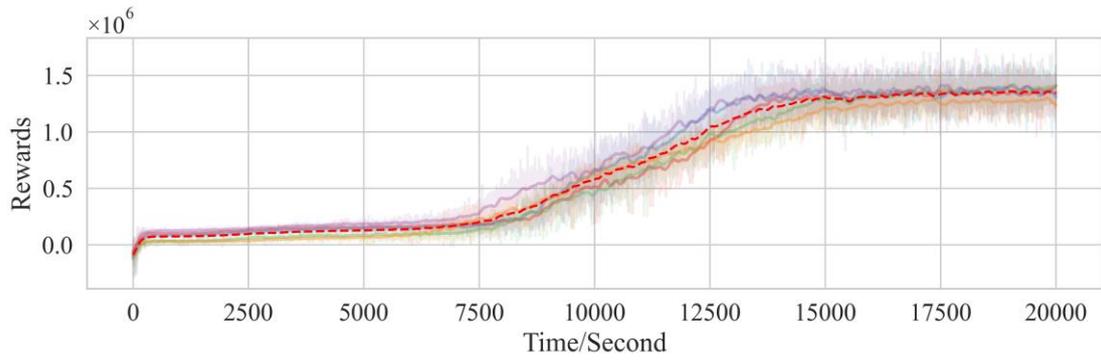

**Fig. 11.** Total Rewards as the Optimization Target in the Training Process.

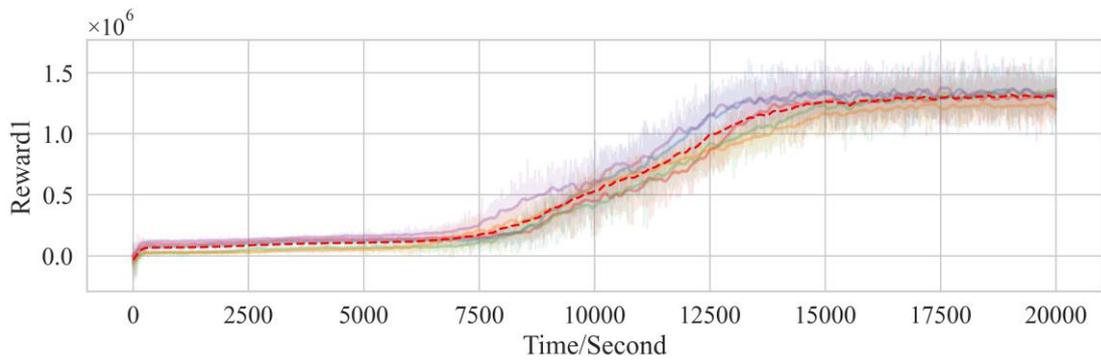

**Fig. 12.** Reward 1 related to Object 1 in the Training Process.

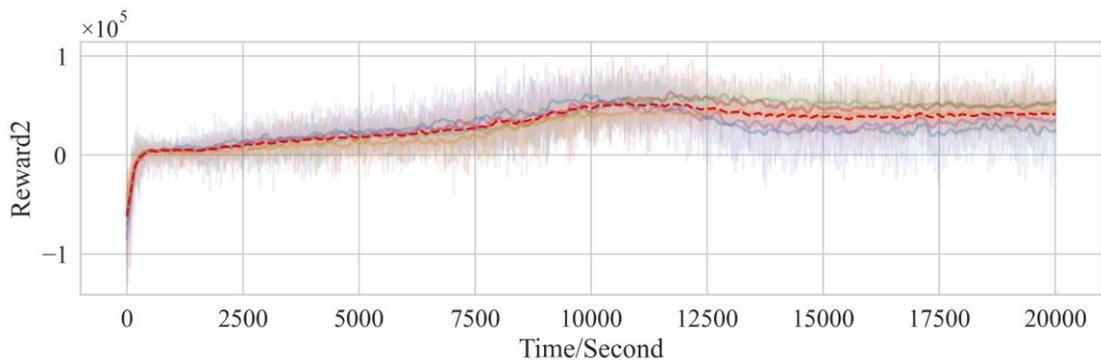

**Fig. 13.** Reward 2 related to Object 2 in the Training Process.

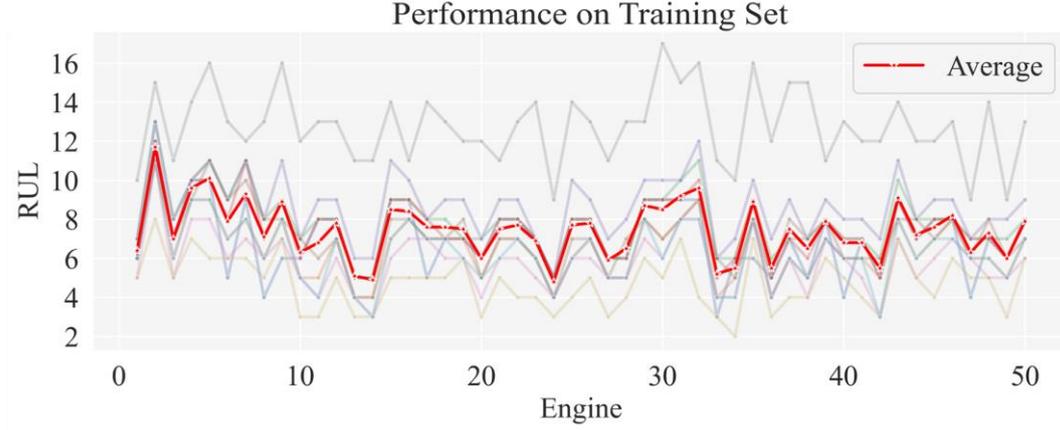

**Fig. 14.** Performance of SMOMA-PPO on FD001 Training Dataset.

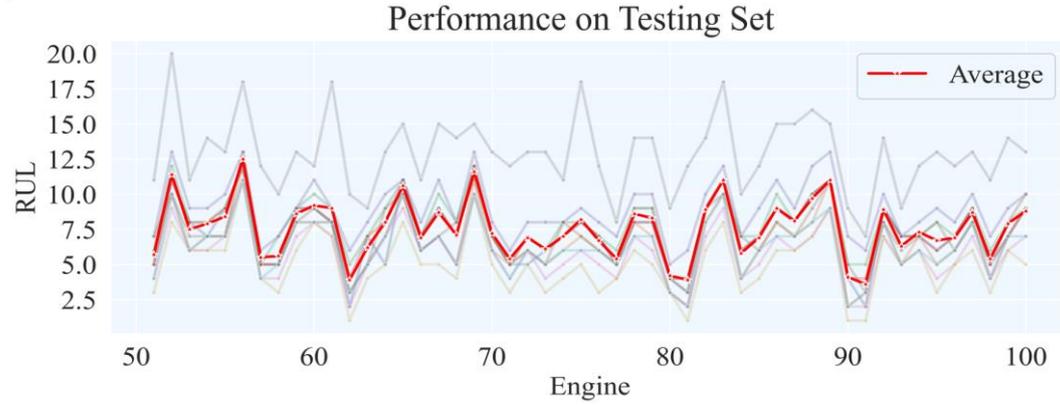

**Fig. 15.** Performance of SMOMA-PPO on FD001 Testing Dataset.

Fig. 14 illustrates the performance of DRL on the training dataset of FD001, where 10 light-colored lines represent 10 random training instances. In contrast, the solid red line is the average of those instances. Fig. 15 demonstrates performance on FD001 testing dataset and the lines of the same color on both Fig. 14 and Fig. 15 represent the performance of the identical trained SMOMA-PPO models. It is evident that when model performs well on the training dataset, it also performs well on the testing dataset. For instance, the gray line in Fig. 14 is positioned at the top, while the tan line is at the bottom. This pattern is similarly observed in Fig. 15.

5.6. Performance of SMOMA-PPO

After calculating the metrics mentioned in section 5.4 for 10 random experiments, the average and standard deviation of these metrics were calculated as shown in Table 6. The related studies with similar research objectives [23] we compared only demonstrated model performance on FD002 and obtained an average RUL of 12.81. In contrast, our model achieved an average of 8.93, only 6.03% of the average engine life in FD002 (19231/130≈147.93, 8.93/147.93≈6.03%). Most importantly, over 95% of the engines from FD002 were replaced when their RUL was less than 20, whereas the compared methods achieved 82% [23]. We also created box plots for the first four metrics to provide a more intuitive understanding, as shown in Fig. 16. Refer to Section 5.3 for the meanings of the abbreviations in the header.

**Table 6.** Performance of SMOMA-PPO on All C-MAPSS Dataset

| Dataset | MR | SR | MAR | MIR | SER | URE | WRE |
|---|---|---|---|---|---|---|---|
| FD001 | 7.29 | 2.15 | 12.20 | 2.90 | 49.60±0.66 | 0.40±0.66 | 0 |
| FD002 | 8.93 | 3.80 | 18.50 | 2.50 | 129.50±0.67 | 0.50±0.67 | 0 |
| FD003 | 7.14 | 2.31 | 12.30 | 2.30 | 49.90±0.30 | 0.10±0.30 | 0 |
| FD004 | 10.89 | 4.55 | 23.00 | 2.00 | 120.00±0.44 | 0 | 4.00±0.44 |

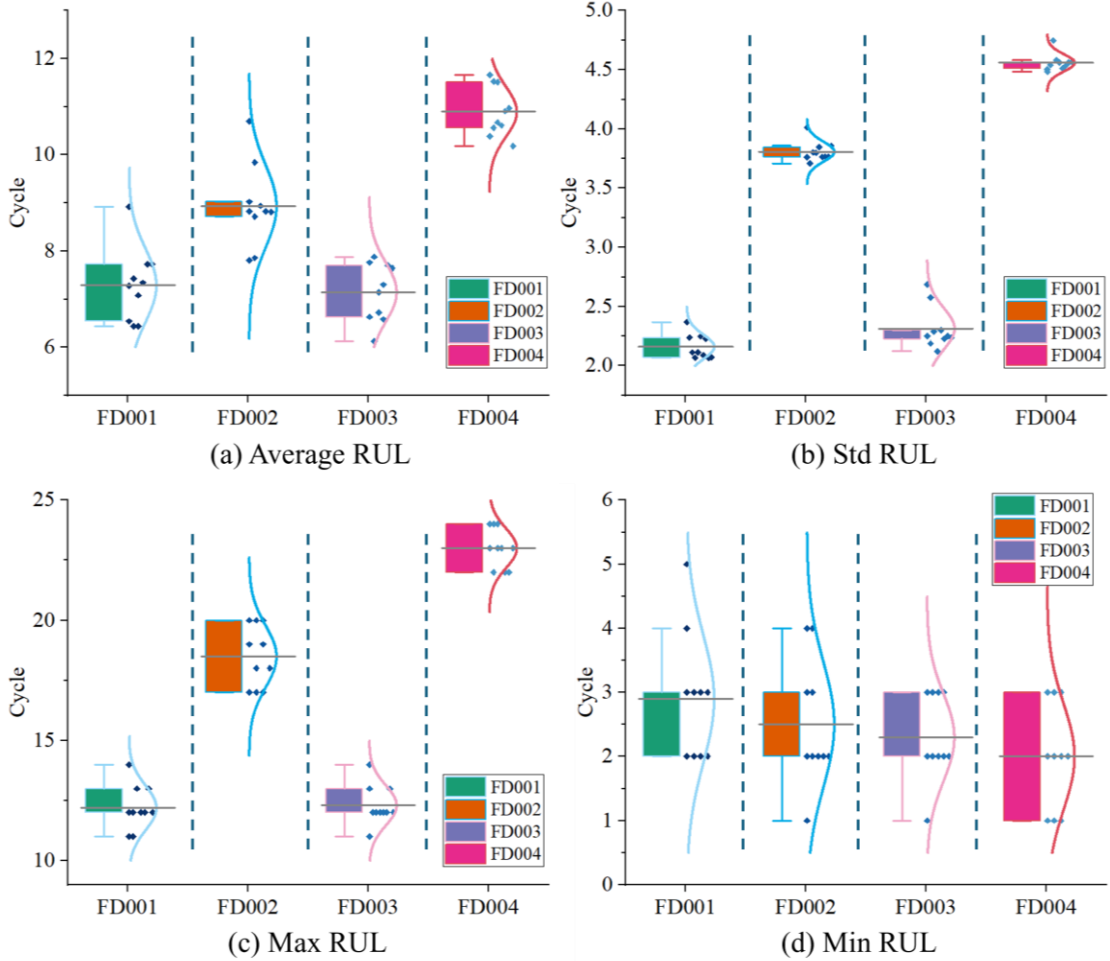

**Fig. 16.** Performances of the model achieved in subsets.

The overall metrics for FD002 and FD004 are higher than those of the other two subsets. This is primarily due to increased data complexity in FD002 and FD004, where the engines have 6 different operation conditions. Additionally, as shown in Table 6, only FD004 exhibits WRE, indicating a small number of engines were replaced when their RUL exceeded 30. This can be attributed to the fact that FD004 has two fault modes, making it the most complex. But the majorities from FD004 were replaced with a low RUL, and there was no UR. Among four subsets, FD003 performed the best. Despite having 2 fault modes, which adds complexity compared to FD001, its performance was not inferior to FD001.

### 5.7. Comparison Among Different Algorithms

To assess and contrast the efficacy of various PdM strategies, we have chosen to evaluate advanced RL-based PdM methodologies and RUL prediction models, namely Prob-RL [23], HIRL [22], AttnPINN [43], and GCU-Trans [40]. Among these, Prob-RL and HIRL represent advanced RL-based approaches. However, no existing methods employ sequential decision-making mechanism. These methods typically rely on single-stage policies where the replacement decision is made directly by a single agent based on the current observation. By contrast, our proposed multi-agent framework introduces a novel sequential resolution mechanism, which addresses the key limitation of single-stage RL approaches, i.e., their inability to balance inspection frequency and replacement timing. On the other hand, given the limited availability of RL-based PdM techniques, we further investigate the disparities between single-agent and multi-agent approaches by employing the PPO and Deep Deterministic Policy Gradient (DDPG) algorithms for this PdM task.

Additionally, as our proposed GRP falls within the realm of RUL prediction methodologies, we include it in this comparative analysis. However, in the realm of RUL prediction approaches, a crucial distinction arises when applying them to PdM tasks. A predetermined threshold must be established. When the predicted RUL falls below this threshold, a replacement action is triggered. This divergence from RL methods represents a significant drawback. To ensure a fair comparison among RUL prediction techniques, we initially conduct experiments to determine a threshold for the GRP method that prevents the occurrence of UR. Subsequently, we apply this identified threshold to other RUL prediction models for identical PdM tasks. The process of identifying an appropriate threshold for the GRP method is detailed in Table 7. Sequentially varying the threshold from 4 to 15, the second column of the table indicates the frequency of UR occurrences at current threshold. The subsequent columns represent the intervals within which the actual RUL of the engine falls upon replacement, once the model reaches the threshold. The final column denotes the average actual RUL of the replaced engines when employing this threshold for PdM tasks. Notably, at a threshold of 12, the model ceases to produce UR, prompting us to use this threshold for comparing different RUL prediction methods. Furthermore, we also evaluate their performances at a threshold of 6, half of 12. Models using the threshold of 12 will be denoted as "-12," while those using 6 will be marked as "-6" for clarity in subsequent discussions.

**Table 7.** Finding the appropriate threshold of GRP to fairly compare the effectiveness of using RUL prediction methods for conducting PdM task.

| Threshold | UR | (0,5] | (5,10] | (10,20] | (20,125] | Average RUL |
|---|---|---|---|---|---|---|
| 4 | 18 | 63 | 40 | 9 | 0 | 5.6250 |
| 5 | 10 | 45 | 59 | 16 | 0 | 6.8166 |
| 6 | 4 | 24 | 73 | 28 | 1 | 8.0158 |
| 7 | 2 | 15 | 71 | 40 | 2 | 9.2578 |
| 8 | 2 | 13 | 55 | 58 | 2 | 10.3906 |
| 9 | 2 | 8 | 46 | 70 | 4 | 11.5546 |
| 10 | 2 | 5 | 38 | 78 | 7 | 12.5000 |
| 11 | 1 | 3 | 32 | 85 | 9 | 13.3953 |
| **12** | **0** | **1** | **20** | **97** | **12** | **14.3384** |
| 13 | 0 | 1 | 13 | 99 | 17 | 15.5230 |
| 14 | 0 | 0 | 8 | 97 | 25 | 17.0000 |
| 15 | 0 | 0 | 7 | 87 | 36 | 18.0077 |

Next comes the performance comparison of all PdM methods, including both RL approaches and RUL prediction methods. We have utilized Ideal Maintenance and Corrective Maintenance as baselines. Ideal maintenance (at true RUL) refers to a scenario where the true RUL is pre-known by an Oracle, enabling engines to be replaced precisely at this predetermined RUL. This strategy involves no unscheduled maintenance tasks and consistently maintains zero wasted engine lives, representing an optimal maintenance approach. Corrective Maintenance, on the other hand, involves replacing engines immediately upon failure, leading to a constant flow of unscheduled replacements, an unfavorable scenario.

Within the Table 8, SRE denotes the successful replacement executions, while UR signifies the occurrences of unscheduled replacements, with the intervals in the table headers carrying the same implications as Table 7. Following the conventional practice in PdM research, we have also assessed the cost associated with each method. The cost computation is outlined as follows: when a replacement is executed with the engine's true RUL falling within (0,5], the cost is minimal, assigned as 1; within (5,10], the cost escalates slightly to 2; within (10,29], the cost increases further due to the engine possessing a considerable RUL, assigned as 3; and within (20,125], the cost peaks at 10. In cases of UR, the cost incurs the maximum penalty of 20. Additionally, we have calculated the average RUL. Furthermore, considering the Inspection Period for PdM methods, which is a relatively scarce metric in existing researches, our focus has primarily centered on comparing against Prob-RL.

**Table 8.** Performance comparison of all PdM methods.

| Approach | | SRE | UR | (0,5] | (5,10] | (10,20] | (20,125] | Cost | Average RUL | Inspection Period |
|---|---|---|---|---|---|---|---|---|---|---|
| | Ideal Maintenance | 130 | 0(0%) | 130 | 0 | 0 | 0 | 130(0%) | / | / |
| | Corrective Maintenance | 0 | 130(100%) | 0 | 0 | 0 | 0 | 2600(100%) | / | / |
| RL | SMOMA-PPO | 130 | 0(0.00%) | 34 | 45 | 51 | 0 | 277(5.95%) | 8.4240 | 32.54 |
| | Prob-RL | 128 | 2(1.54%) | 36 | 42 | 49 | 1 | 317(7.57%) | 8.9692 | 30.00 |
| | HIRL | 127 | 3(2.31%) | 34 | 55 | 38 | 0 | 318(7.61%) | 9.3437 | 1.00 |
| | SAPPO | 125 | 5(3.85%) | 27 | 53 | 45 | 0 | 368(9.64%) | 10.1200 | 1.00 |
| | SADDPG | 126 | 4(3.08%) | 26 | 54 | 43 | 3 | 373(9.84%) | 9.3015 | 1.00 |
| DL | GRP-12 | 130 | 0(0.00%) | 1 | 20 | 97 | 12 | 452(13.04%) | 14.3384 | 1.00 |
| | GRP-6 | 126 | 4(3.08%) | 26 | 71 | 28 | 1 | 342(8.58%) | 8.0158 | 1.00 |
| | AttnPINN-12 | 129 | 1(0.77%) | 2 | 46 | 76 | 5 | 392(10.61%) | 11.7984 | 1.00 |
| | AttnPINN-6 | 110 | 20(15.38%) | 59 | 41 | 10 | 0 | 571(17.85%) | 5.3636 | 1.00 |
| | GCU-Trans-12 | 129 | 1(0.77%) | 5 | 38 | 80 | 6 | 401(10.97%) | 12.0465 | 1.00 |
| | GCU-Trans-6 | 113 | 17(13.08%) | 56 | 49 | 8 | 0 | 518(15.71%) | 5.6106 | 1.00 |

Except SMOMA-PPO, all existing RL-based methods have only 1 agent, facing inherent inability of adjusting the inspection, while our method dynamically handles this based on the prediction of agent 2, which has brought a significant improvement. For instance, Prob-RL incurs a higher cost (317) and 2 URs with a fixed inspection interval (30 cycles). HIRL slightly increases UR to 3 and maintains a comparable cost (318). Other single-agent RL methods, SAPPO and SADDPG, exhibit even higher UR occurrences, reflecting inefficiencies in both inspection and replacement decisions. In contrast, SMOMA-PPO eliminates all URs and achieves the lowest cost (277). Additionally, it achieves an average inspection period of 32.54 cycles, surpassing Prob-RL's fixed interval (30) and significantly outperforming other approaches, which rely on per-cycle inspections. These results clearly demonstrate the robustness and efficiency of our multi-agent approach, which not only reduces costs but also enhances system reliability by dynamically adapting inspection and replacement schedules. This improvement underscores the unique advantage of employing multiple agents to address the real-world predictive maintenance tasks.

In addition to RL-based methods, the RUL prediction-based approaches exhibit noticeable limitations in this PdM task, as reflected in Table 8. For instance, GRP-12 achieves no URs, but its cost is significantly higher (452) due to excessive replacements within the range of (20,125] RUL, highlighting its inefficiency in balancing replacement timing. GRP-6, with a lower threshold, incurs 4 URs and a reduced cost (342), but still underperforms compared to RL-based approaches like Prob-RL, as it heavily relies a pre-defined rigid threshold. Other methods, such as AttnPINN and GCU-Trans, demonstrate similar drawbacks. AttnPINN-12 and GCU-Trans-12 achieve 1 UR each, but their costs remain high at 392 and 401, respectively, primarily due to frequent premature replacements. When the threshold is reduced to 6, both AttnPINN-6 and GCU-Trans-6 exhibit a dramatic increase in UR occurrences (20 and 17), leading to considerably high costs (571 and 518).

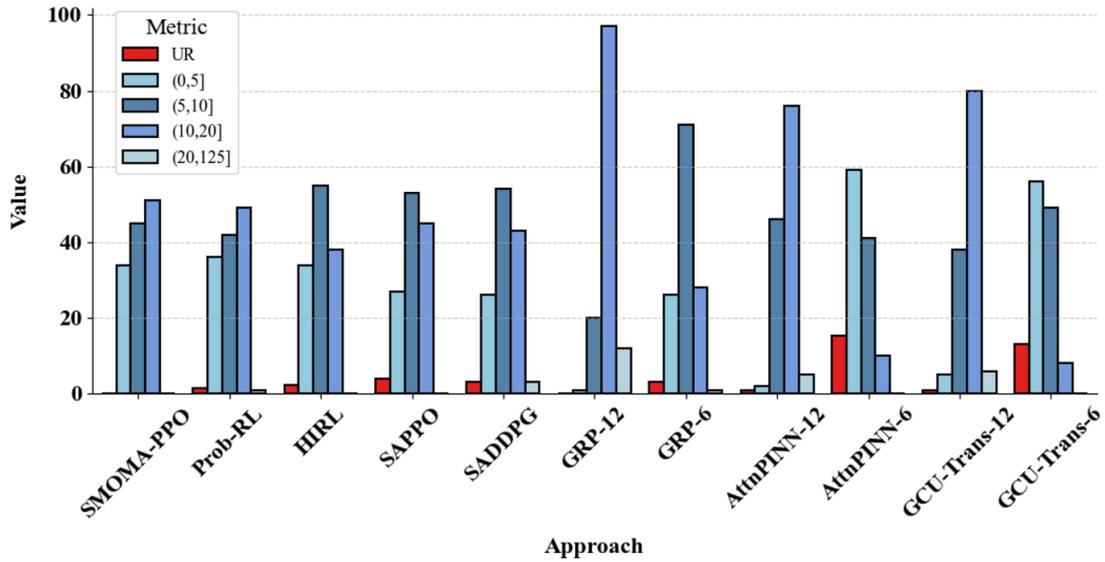

**Fig. 17.** Frequency distribution showing the number of replacements each method performs across different RUL intervals.

To further illustrate the comparison, we visualized the distribution of all replacement actions across different RUL intervals for each approach as shown in Fig. 17, which highlights distinct patterns among the methods. SMOMA-PPO demonstrates a balanced distribution, with most replacements occurring in the (5,10] and (10,20] ranges, reflecting its ability to optimize replacement timing while avoiding URs. Other RL-based methods, such as Prob-RL and HIRL, show a similar trend but with higher occurrences of replacements in near-failure interval (0,5] or beyond (20,125], indicating less efficient decision-making. In contrast, DL-based methods show more scattered distributions, as seen with GRP-12 and GCU-Trans-12. When thresholds are reduced (e.g., GRP-6 and AttnPINN-6), replacements shift towards (0,5], leading to increased occurrences of URs. These patterns emphasize the less adaptive nature of DL-based approaches.

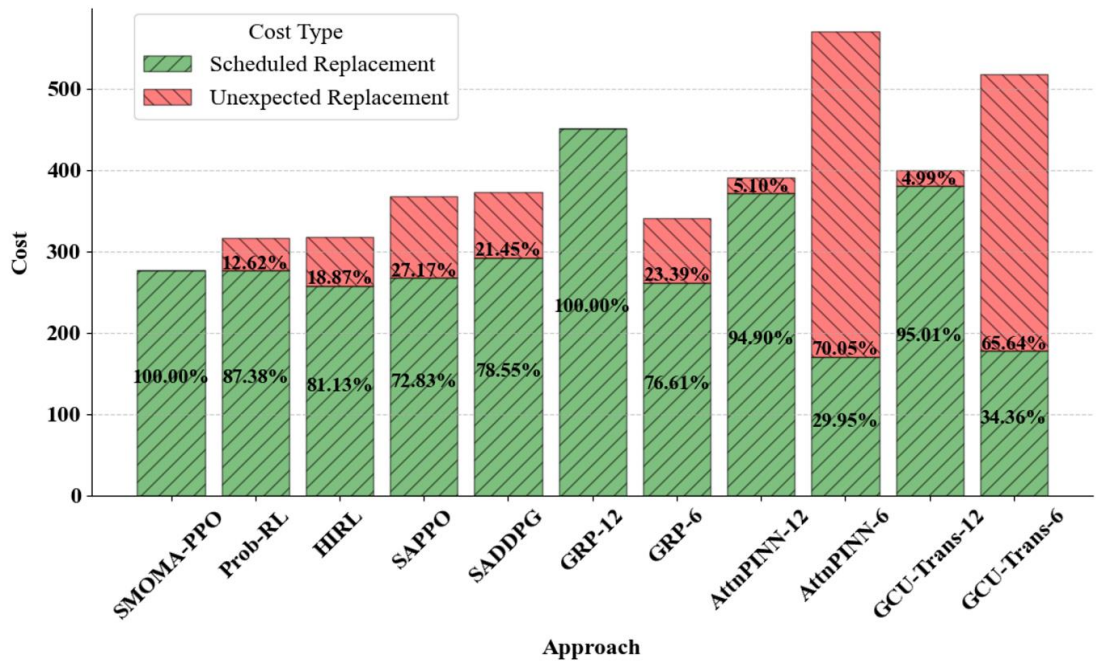

**Fig. 18.** Stacked bar chart separating the costs caused by SRE and UR for each method.

Fig. 18 illustrates the breakdown of costs for each method, separating the contributions of SRE

(green) and UR (red). As shown, our proposed method, SMOMA-PPO, achieves the lowest total cost, entirely attributed to scheduled replacements, with no cost from URs. In contrast, other RL-based methods incur higher costs due to their inability to fully eliminate unscheduled events. DL-based methods, particularly GRP-12, exhibit high scheduled replacement costs due to premature replacements, while methods like AttnPINN-6 and GCU-Trans-6 suffer from significant UR costs. This visualization highlights the effectiveness of SMOMA-PPO in maintaining a balanced and cost-efficient maintenance strategy.

At last, we have also expressed the position of each UR and Cost value in percentage terms relative to Ideal Maintenance and Corrective Maintenance. Finally, for a more visual representation of the algorithms' functionality, we sampled eight engines from the FD002 testing dataset to demonstrate the specific PdM processes of models trained on the training dataset as Fig. 19. Taking engine 131 as an example, in the illustration, the blue rectangular boxes represent the actions of agent 1, while the purple dialog boxes represent the actions of agent 2. At the initial time point, the cycle count has already reached 60, as in the GRP algorithm, the sliding window length is set to 60. At the starting point, the first agent determines that the engine should not be replaced at the current moment, hence the action provided is "continue." The second agent, based on the current environmental state, which was generated by employing the GRP algorithm and the cumulative probability values obtained from function fitting over the last 10 cycles, to decide the next appropriate inspection time. Agent 2 suggests an action value of 43, which shifts the next assessment point forward by 43 cycles, and this process continues accordingly. This iterative process continues until the true RUL of the engine reaches 8. At this point, the first agent concludes that the engine should be replaced, executing the replacement action as indicated by the final yellow box.

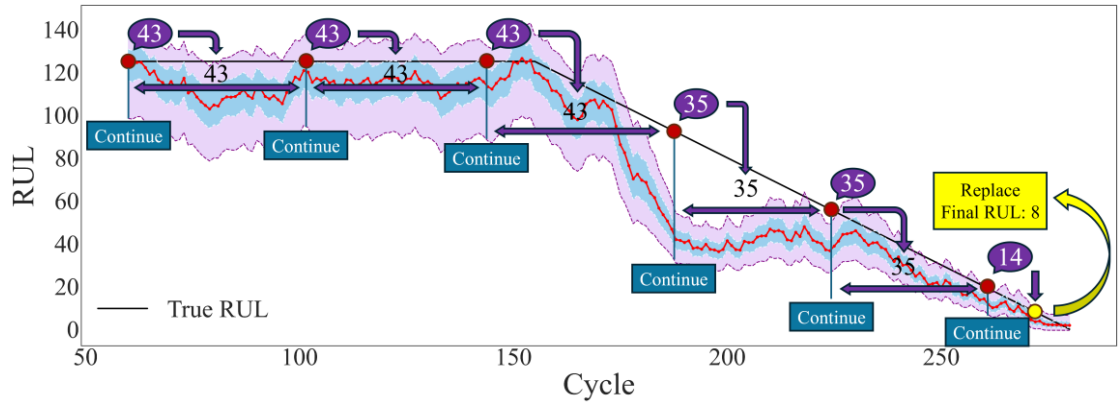

(a) Engine 131

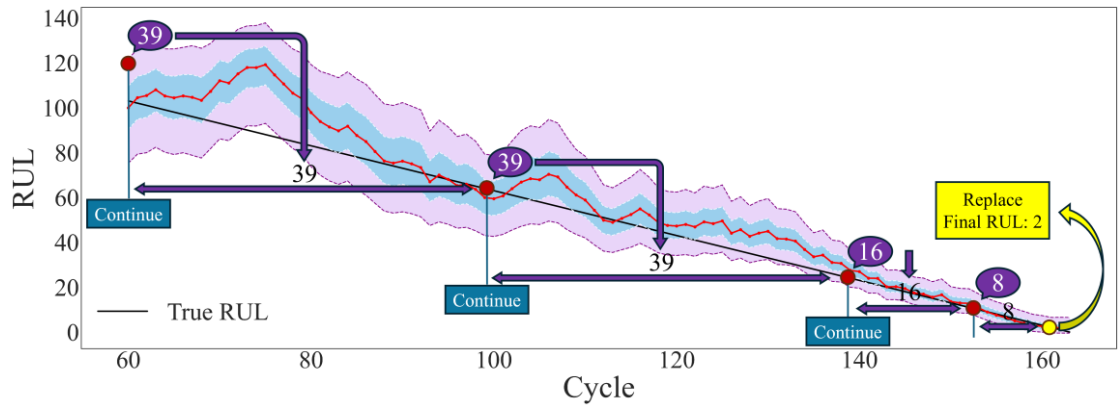

(b) Engine 141

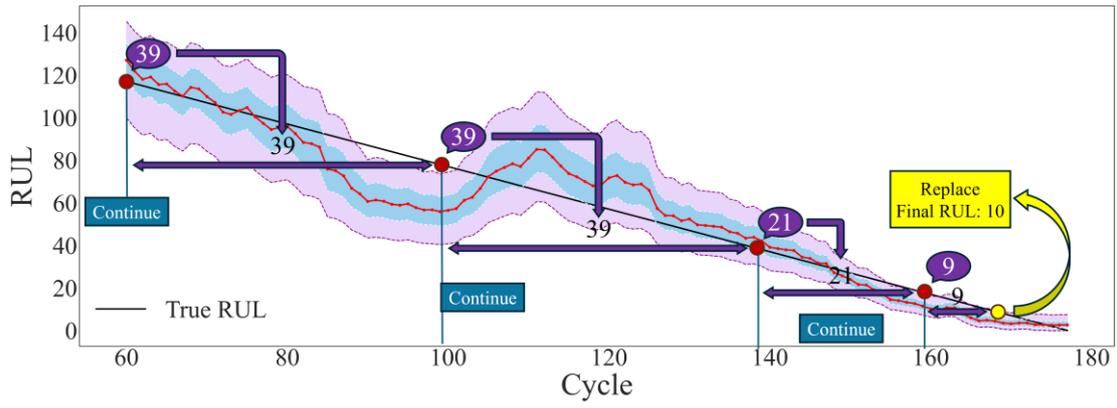

(c) Engine 151

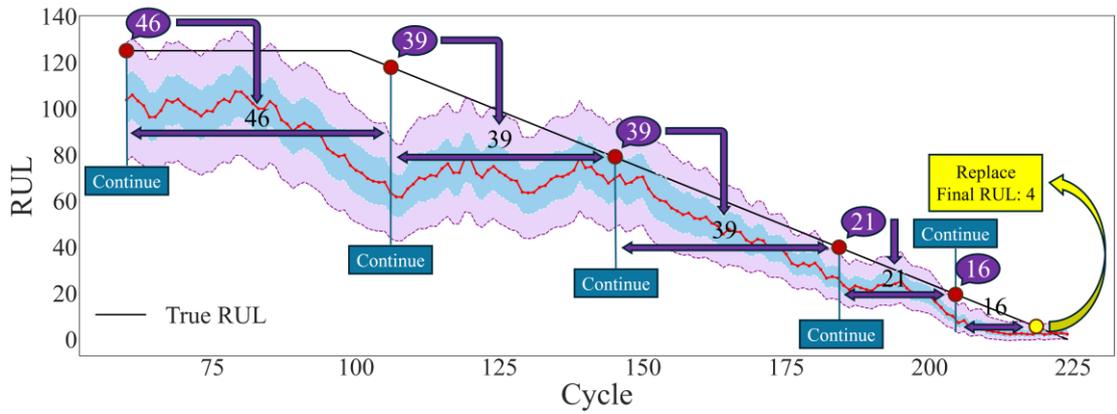

(d) Engine 161

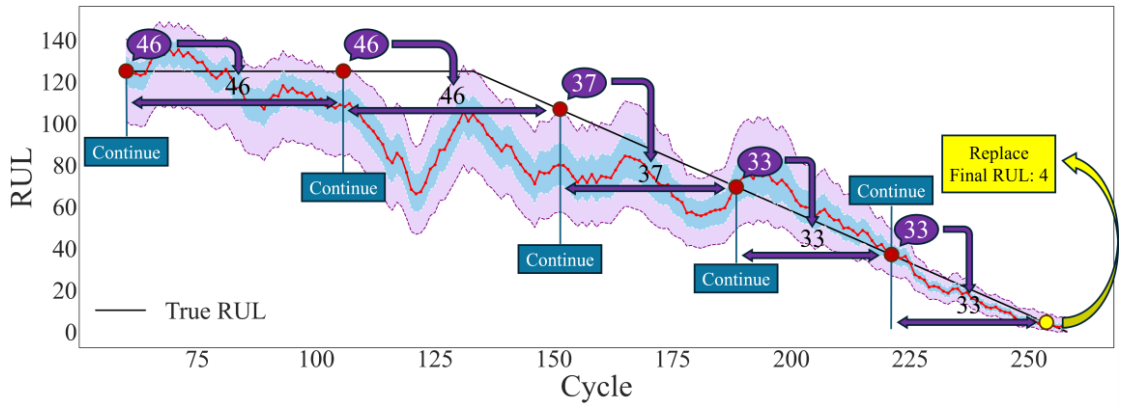

(e) Engine 171

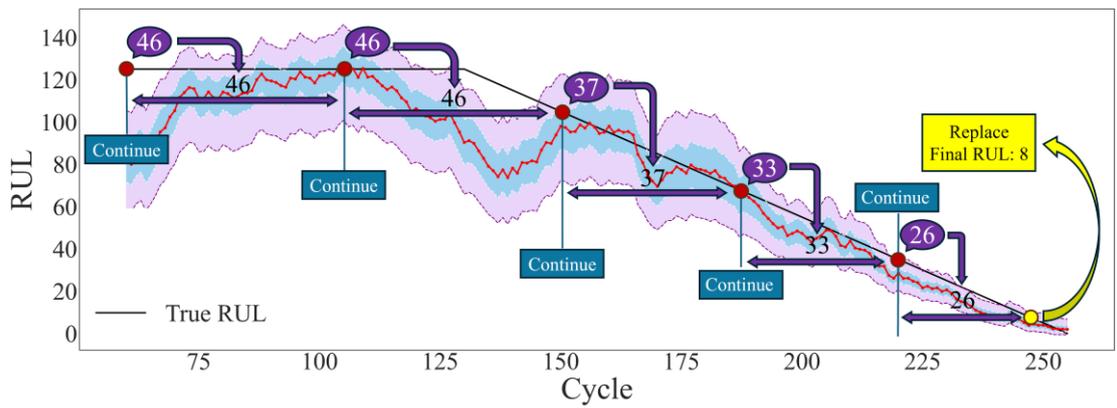

(f) Engine 181

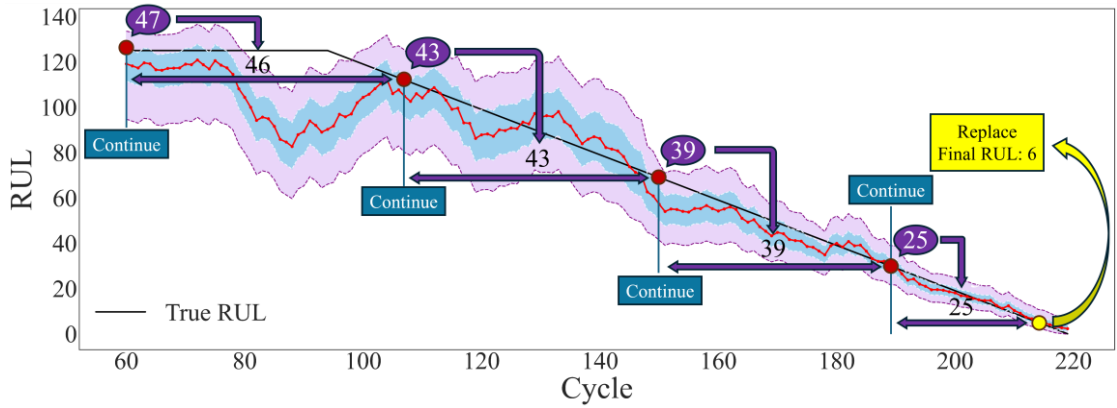

(g) Engine 191

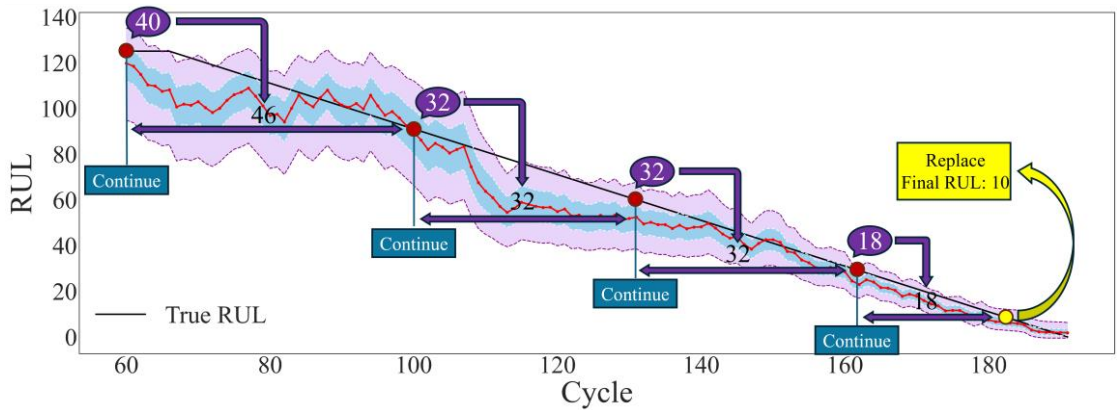

(h) Engine 201

**Fig. 19.** The detailed PdM processes of SMOMA-PPO model trained on the FD002 training dataset.

6. Conclusion

To tackle the challenges of adaptive PdM considering multiple objectives for turbofan engines, this study has introduced a novel approach harnessing MARL to handle the complex, multi-objective nature of PdM tasks. The innovative SMOMA-PPO framework not only aims to minimize the RUL of replaced engines but also autonomously determines optimal inspection periods to mitigate costs.

Furthermore, to equip the RL agents with decision-making insights, we have devised a novel Probabilistic RUL Prediction model based on QR, which provides a distribution of RUL estimates and successfully integrates uncertainty into the decision-making process of the RL agent. The inclusion of uncertainty not only significantly enhanced the model's robustness but also enabled the agent to dynamically adjust maintenance strategies, thus ensuring SMOMA-PPO's performance.

After extensive experimentation and evaluation, SMOMA-PPO has surpassed existing methodologies. It has significantly improved average RUL outcomes compared to current research with similar goals, while simultaneously reducing URs. Moreover, in a domain with limited exploration regarding decreasing inspection frequencies, our model has excelled by achieving the lengthiest average inspection period. The RL agent adeptly adjusts inspection intervals as operational cycles progress, resulting in notable reductions in labor and material costs.

Looking ahead, numerous promising avenues for further research emerge. Firstly, integrating additional contextual information, such as environmental parameters, could bolster the precision and robustness of the PdM model. Secondly, deploying our model in more complex industrial contexts holds substantial promise.

Fundamentally, our model provides a practical and efficient solution for industrial PdM applications characterized by multiple intricate objectives involving temporal sequencing and interdependencies, contributing to the advancement of AI applications in industrial settings and paving the way for further enhancements in engineering reliability.

**CRediT authorship contribution statement**

Yan Chen: Conceptualization, Methodology, Software, Writing – original draft. Cheng Liu: Writing – review & editing, Supervision, Funding acquisition.

**Declaration of competing interest**

The authors declare that they have no known competing financial interests or personal relationships that could have appeared to influence the work reported in this paper.

**Data availability**

The datasets come from a public database.